\documentclass[floatfix,twocolumn,showpacs,preprintnumbers,amsmath,amssymb,pre,superscriptaddress]{revtex4-1}
\usepackage{color}
\usepackage[usenames,dvipsnames,svgnames,table]{xcolor}
\usepackage[colorlinks=true,linkcolor=blue,urlcolor=blue,citecolor=blue]{hyperref}

\usepackage{graphicx}
\usepackage{dcolumn}
\usepackage{bm}
\usepackage{subfigure}
\usepackage{amssymb}
\usepackage{multirow}
\usepackage{amsmath}
\usepackage{braket}
\graphicspath{{plots/}}

\begin{document}

\title{\textit{Ab Initio} Path Integral Monte Carlo Approach\\ to the Static and Dynamic Density Response of the Uniform Electron Gas}


\author{S.~Groth}

\affiliation{Institut f\"ur Theoretische Physik und Astrophysik, Christian-Albrechts-Universit\"at zu Kiel,
 Leibnizstra{\ss}e 15, D-24098 Kiel, Germany}

\author{T.~Dornheim}
\email{t.dornheim@hzdr.de}

\affiliation{Center for Advanced Systems Understanding, Untermarkt 20, D-02826 G\"orlitz, Germany}

\affiliation{Institut f\"ur Theoretische Physik und Astrophysik, Christian-Albrechts-Universit\"at zu Kiel,
 Leibnizstra{\ss}e 15, D-24098 Kiel, Germany}

\author{J.~Vorberger}

\affiliation{Helmholtz-Zentrum Dresden-Rossendorf, Bautzner Landstra{\ss}e 400, D-01328 Dresden, Germany}



\begin{abstract}
In a recent Letter~[T.~Dornheim \textit{et al.}, \textit{Phys.~Rev.~Lett.}~\textbf{121}, 255001 (2018)] we have presented the first \textit{ab initio} results for the dynamic structure factor $S(\mathbf{q},\omega)$ of the uniform electron gas for conditions ranging from the warm dense matter regime to the strongly correlated electron liquid. This was achieved on the basis of exact path integral Monte Carlo data by stochastically sampling the dynamic local field correction $G(\mathbf{q},\omega)$. In this paper, we introduce in detail
this new reconstruction method and provide several practical demonstrations. Moreover, we thoroughly investigate the associated imaginary-time density--density correlation function $F(\mathbf{q},\tau)$. The latter also gives us access to the static density-response function $\chi(\mathbf{q})$ and static local field correction $G(\mathbf{q})$, which are compared to standard dielectric theories like the widespread random phase approximation. In addition, we study the high-frequency limit of $G(\mathbf{q},\omega)$ and provide extensive new results for the dynamic structure factor for different densities and temperatures. Finally, we discuss the implications of our findings for warm dense matter research and the interpretation of experiments.

\end{abstract}

\maketitle

\section{Introduction}

The uniform electron gas (UEG) is one of the most important model systems in physics and quantum chemistry~\cite{quantum_theory,loos}. First and foremost, it was only the accurate parametrization of the exchange-correlation energy of the UEG~\cite{vwn,perdew} based on zero-temperature quantum Monte Carlo (QMC) calculations~\cite{gs1,gs2} that facilitated the spectacular success of density functional theory simulations of real materials~\cite{dft_review}. Moreover, the UEG has been of paramount importance for several breakthroughs in theoretical physics such as Fermi liquid theory~\cite{pines} or the Bardeen-Cooper-Schrieffer theory of superconductivity~\cite{bcs,bcs2}.

While most static properties of the UEG have been known at zero temperature for decades, there recently has emerged a growing interest into the properties of electrons at extreme densities and temperatures~\cite{ross,koenig,fortov_review}. In fact, this so-called warm dense matter (WDM) regime constitutes one of the most active frontiers in plasma physics and is of fundamental importance for, e.g., the description of astrophysical objects like giant planet interiors~\cite{militzer1,militzer2,knudson,militzer3,manuel} and brown dwarfs~\cite{saumon1,saumon2,becker}, laser-excited solids~\cite{ernstorfer2,ernstorfer}, and the pathway towards inertial confinement fusion~\cite{nora,schmit,hurricane3,kritcher}---the latter potentially offering a near abundance of clean energy in the future.
In addition, WDM is now routinely realized in experiments with, e.g., free-electron lasers~\cite{lcls1,sperling,xfel1} or diamond anvil cells~\cite{benuzzi} in large research facilities around the globe, see Ref.~\cite{falk_wdm} for a topical overview.

From a theoretical perspective, the WDM regime is defined by two parameters that are both of the order of unity: 1) the density parameter (sometimes denoted as \emph{quantum coupling parameter} or Wigner-Seitz-radius) $r_s=\overline{r}/a_\textnormal{B}$, with $\overline{r}$  and $a_\textnormal{B}$ being the average inter-particle distance and Bohr radius, and 2) the degeneracy temperature $\theta=k_\textnormal{B}T/E_\textnormal{F}$, with $E_\textnormal{F}$ being the Fermi energy. Strictly speaking, a third parameter is given by the classical coupling parameter $\Gamma=Z e^2/(\overline{r}k_\textnormal{B}T)\sim1$ describing the ionic component in real WDM systems, but it is of no relevance for the UEG and, therefore, is not further discussed in the present work. 

Speaking in terms of physical effects, WDM is characterized by the intriguingly intricate interplay of the Coulomb repulsion between the electrons with thermal excitations and quantum degeneracy effects, such as quantum diffraction and Pauli blocking. Moreover, there are no small parameters to conduct an expansion around, which leaves numerical methods as the only option, with QMC approaches being particularly promising~\cite{wdm_book}. This renders WDM theory a notoriously tricky business, as QMC simulations of electrons are severely hampered by the infamous fermion sign problem (FSP)~\cite{dornheim_pop,troyer,loh}. 
Consequently, despite intensive efforts regarding the warm dense UEG over several decades~\cite{pines,kugler1,F,ebeling,stls,rajagopal,pdw,stolzmann,sandipan,brown_ethan,vladimir_UEG,dornheim_test}, the accurate description of this system has only been achieved recently. In particular, Groth, Dornheim and co-workers~\cite{groth_prl,review} have presented an accurate parametrization of the exchange-correlation free energy $f_\textnormal{XC}$ of the warm dense UEG on the basis of different path integral Monte Carlo (PIMC) techniques~\cite{dornheim,groth,dornheim2,dornheim3,dornheim_prl,dornheim_neu}, which is available over the entire relevant parameter range and provides a full thermodynamic description, see Ref.~\cite{review} for a topical review article.

In spite of these significant advances, one crucial piece of the bigger puzzle that is the theory of the warm dense electron gas is yet missing: the response of an electron gas to an external perturbation. In particular, the experimental observation and subsequent theoretical description of the time-dependent density response is of paramount importance as a method of diagnostics in modern WDM applications~\cite{siegfried_review,valenzuela,dominik}.
In this context, the central quantity is given by the dynamic density--density response function $\chi(\mathbf{q},\omega)$, with $\mathbf{q}$ and $\omega$ being the wave vector and frequency, or, equivalently, the dynamic structure factor $S(\mathbf{q},\omega)$, that is directly measured in X-ray Thomson scattering experiments, see Ref.~\cite{siegfried_review} for a review. 
In addition to its utility for the interpretation of experiments, the dynamic density-response is important as input for many nonequilibrium applications such as the stopping power~\cite{Cayzac:2017,Fu:2017}, energy transfer rates and relaxation~\cite{transfer1,transfer2}, electrical and thermal conductivities~\cite{Desjarlais:2017,Veysman:2016}, quantum hydrodynamics~\cite{zhandos}, or the development of advanced exchange--correlation functionals for density function theory~\cite{lu,patrick,burke2}.

Unfortunately, an exact theory for $\chi(\mathbf{q},\omega)$ is even more challenging than for the previously discussed static properties, as one, in principle, would have to carry out an explicit propagation in time. Needless to say, this is in general not possible for an interacting quantum many-body system except in a few limiting cases. Consequently, the dynamic density response is relatively poorly understood even in the ground state, see Refs.~\cite{takada1,takada2} for the presumably most accurate data. Moreover, \textit{ab initio} QMC methods are limited to the description of the static limit~\cite{bowen,moroni} [i.e., a constant, time-independent external perturbation described by $\chi(\mathbf{q})=\chi(\mathbf{q},\omega=0)$], and even this task has turned out to be relatively expensive and QMC data are only available at a few selected parameters~\cite{moroni2,bowen2}. For completeness, we mention that this approach has recently been adapted to the WDM regime in Refs.~\cite{dornheim_pre,groth_jcp}, but here the available data points have remained even more sparse.

On the other hand, it has long been known that PIMC methods~\cite{berne1,berne2} allow for a straightforward calculation of the \emph{imaginary-time}  density--density correlation function $F(\mathbf{q},\tau)$, cf.~Eq.~(\ref{eq:F}) below. In addition to direct access to the static density response, this quantity can be used as input for the \emph{reconstruction} of $S(\mathbf{q},\omega)$, which is a well known, but notoriously difficult problem~\cite{jarrell,schoett}. In particular, the obtained solution for $S(\mathbf{q},\omega)$ are often not unique since the PIMC data for $F$ are afflicted with a statistical uncertainty, and one somehow has to provide additional input to render the reconstruction tractable. Indeed, we have recently~\cite{dornheim_dynamic} proposed a novel reconstruction procedure based on the stochastic sampling of the dynamic local field correction (LFC) $G(\mathbf{q},\omega)$, which allows to automatically fulfill a number of exact constraints. This, in turn, has allowed us to present the first accurate data for $S(\mathbf{q},\omega)$ of the UEG going from the WDM regime to the strongly correlated electron liquid.

In this paper, we further explore this new procedure and present in detail the involved theory, provide practical examples, and give extensive new data both for static and dynamic quantities for hitherto unexplored parameters.
In addition to the value of our results for the description of matter under extreme conditions, we expect our new reconstruction procedure to be of broad interest for different communities, such as ultracold bosonic atoms~\cite{dynamic_alex1,dynamic_alex2,vitali,supersolid} or condensed matter physics~\cite{gull,som,igorrr}.

The paper is organised as follows: in Sec.~\ref{sec:theory}, we introduce the required theory, starting with the utilized path integral Monte Carlo method (\ref{sec:PIMC}) and linear response theory (\ref{sec:LRT}). Furthermore, we introduce the concept of imaginary-time correlation-functions and their relation to the dynamic structure factor (\ref{sec:ITCF}), the general problem of reconstruction (\ref{sec:reconstruction_results}), and our new stochastic sampling procedure (\ref{sec:stochastic_sampling}) based on the dynamic LFC.
The discussion of our results in Sec.~\ref{sec:results} starts with an investigation of $F(\mathbf{q},\tau)$ (\ref{sec:F_results}), followed by the static properties of the UEG (\ref{sec:static_response}), and the respective high-frequency ($\omega\to\infty$) limit.
In Sec.~\ref{sec:stochastic_sampling_results}, we present two practical examples of the reconstruction of $S(\mathbf{q},\omega)$ and subsequently (\ref{sec:dynamic_results}) give new results for previously unexplored conditions.
The paper is concluded by a brief summary and discussion in Sec.~\ref{sec:summary}.

We assume Hartree atomic units throughout this work.


\section{Theory\label{sec:theory}}

\subsection{Path Integral Monte Carlo\label{sec:PIMC}}

Let us consider a system of $N$ unpolarized electrons (i.e., with identical numbers of spin-up and -down electrons, $N_\uparrow = N_\downarrow = N/2$) in a volume $V=L^3$ at an inverse temperature $\beta=1/k_\textnormal{B}T$. In this case, all thermodynamic observables can be computed from the canonical partition function, which, in coordinate space, is given by
\begin{eqnarray}\label{eq:Z}
 Z &=& \frac{1}{N_\uparrow!N_\downarrow!} \sum_{\sigma_\uparrow\in S_{N_{\uparrow}}}\sum_{\sigma_\downarrow\in S_{N_{\downarrow}}} \textnormal{sgn}^\textnormal{f}(\sigma_\uparrow,\sigma_\downarrow) \\ \nonumber
 & &
 \int \textnormal{d}\mathbf{R}\ \bra{ \mathbf{R} } e^{-\beta\hat{H}} \ket{ \hat{\pi}_{\sigma_\uparrow}\hat{\pi}_{\sigma_\downarrow}\mathbf{R}} \quad ,
\end{eqnarray}
with $\sigma_i$ denoting a particular element from the permutation group $S_{N_i}$, and $\hat{\pi}_{\sigma_i}$ being the corresponding permutation operator with $i\in\{\uparrow,\downarrow\}$. Observe that in this notation $\mathbf{R}$ contains the coordinates of both spin-up and -down electrons, and the sign function $\textnormal{sgn}^\textnormal{f}(\sigma_\uparrow,\sigma_\downarrow)$ can be positive or negative depending on the number of pair-exchanges. The problem with Eq.~(\ref{eq:Z}) is that the matrix elements of the density operator $\hat\rho = e^{-\beta\hat H}$ cannot be evaluated as the kinetic and potential contributions, $\hat K$ and $\hat V$, do not commute, i.e.,
\begin{eqnarray}\label{eq:primitive}
 e^{-\beta\hat H} =  e^{-\beta\hat K} e^{-\beta\hat V} + \mathcal O\left(\beta^2 \right) \quad .
\end{eqnarray}
The basic idea of the path integral Monte Carlo formalism~\cite{berne3,cep} is to perform a Trotter decomposition~\cite{trotter} and express the partition function as the sum over $P$ sets of particle coordinates, but evaluated at a $P$-times higher temperature,
\begin{eqnarray}\label{eq:Z_PIMC}
 Z &=& \frac{1}{N_\uparrow!N_\downarrow!} \sum_{\sigma_\uparrow\in S_{N_{\uparrow}}}\sum_{\sigma_\downarrow\in S_{N_{\downarrow}}} \textnormal{sgn}^\textnormal{f}(\sigma_\uparrow,\sigma_\downarrow) \\ \nonumber
 & &
 \int \textnormal{d}\mathbf{R}_0 \dots \textnormal{d}\mathbf{R}_{P-1} \prod_{\alpha=0}^{P-1} \bra{ \mathbf{R}_\alpha } e^{-\epsilon\hat{H}} \ket{ \hat\pi_{P-1}\mathbf{R}_{\alpha-1}} \\ \nonumber &=& \int \textnormal{d}\mathbf{X}\ W(\mathbf{X}) \quad ,
\end{eqnarray}
with $\epsilon=\beta/P$ and $\hat\pi_{P-1}$ only having an effect for $\alpha = P-1$. Evidently, the factorization error from Eq.~(\ref{eq:primitive}) can be made arbitrarily small by increasing the convergence parameter $P$, which in turn means that any desired level of accuracy can be realized and the PIMC formalism is quasi-exact. Note that the sign $\textnormal{sgn}^\textnormal{f}(\sigma_\uparrow,\sigma_\downarrow) $ depends on the parity of a particular permutation of particle coordinates and flips for every pair-exchange.
Further, it holds $\mathbf{R}_0=\mathbf{R}_{P-1}$, which implies that Eq.~(\ref{eq:Z_PIMC}) can be interpreted as the sum over all closed paths $\mathbf{X}$ of particle coordinates in the so-called imaginary-time $\tau=-i\hbar\beta$.
\begin{figure}
\includegraphics[width=0.4147\textwidth]{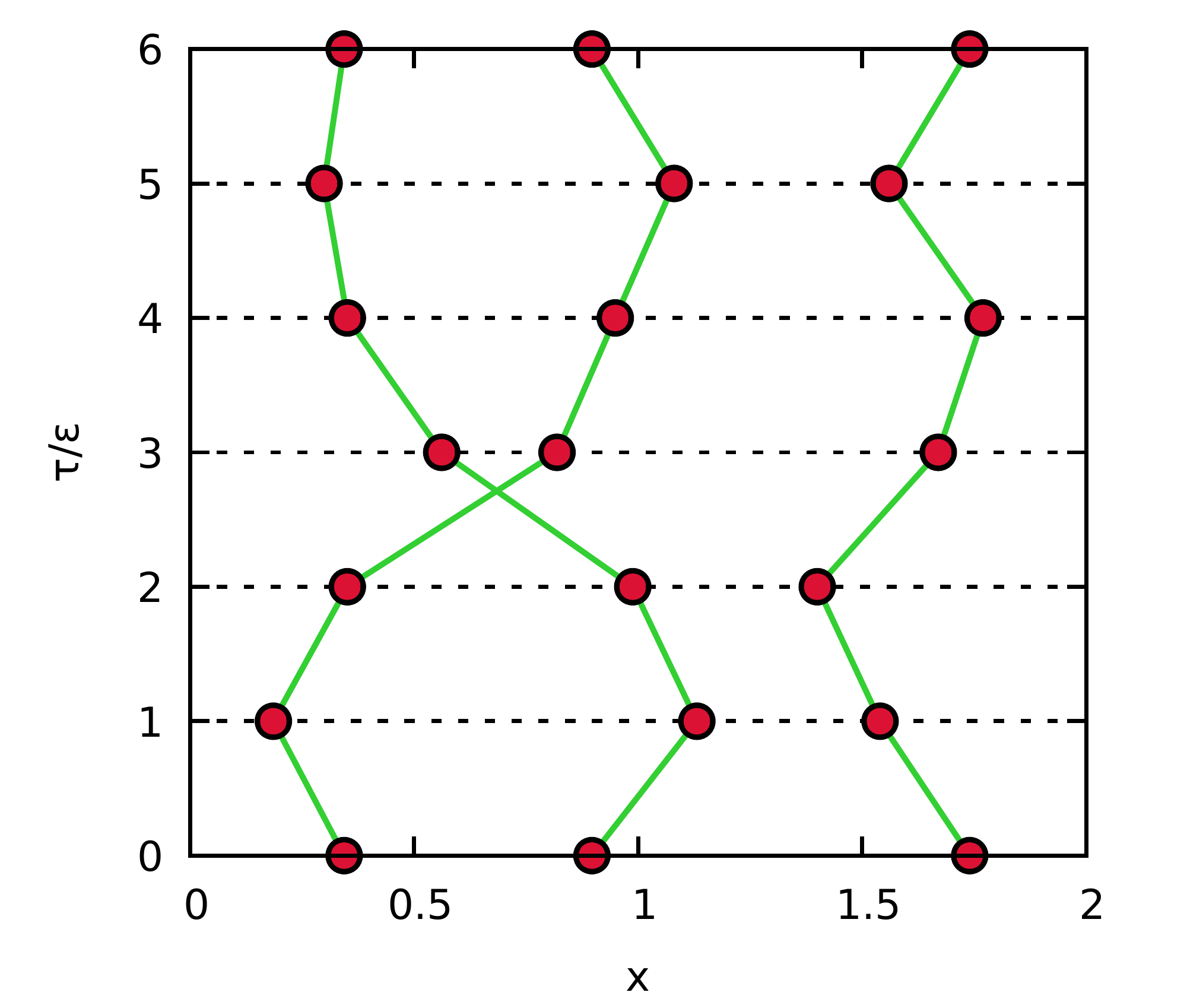}
\caption{\label{fig:PIMC}
Schematic illustration of Path Integral Monte Carlo---Shown is a configuration of $N=3$ electrons with $P=6$ imaginary--time propagators in the $x$-$\tau$ plane. Due to the single pair-exchange, the corresponding configuration weight $W(\mathbf{X})$ is negative.
}
\end{figure}  
This is illustrated in Fig.~\ref{fig:PIMC}, where we show a configuration of $N_\uparrow=3$ electrons in the $x$-$\tau$-plane. In this particular example, we are having $P=6$ imaginary-time slices, with the coordinates at $\alpha=0$ and $\alpha=6$ being identical. In addition, the corresponding configuration weight $W(\mathbf{X})$ is negative due to the presence of a single pair-exchange.
Moreover, each high-temperature factor can be viewed as a propagation in the imaginary time by a time-step $\epsilon$, which allows for straightforward measurements of imaginary-time correlation-functions as discussed in Sec.~\ref{sec:ITCF}.

The basic idea of the PIMC approach is to use the Metropolis algorithm~\cite{metropolis} to stochastically sample the paths $\mathbf{X}$ according to $W(\mathbf{X})$. Unfortunately, however, this is not directly possible as $W(\mathbf{X})$ is not strictly positive and, thus, cannot be interpreted as a probability distribution. To circumvent this issue, we consider the modified partition function
\begin{eqnarray}
Z' = \int  \textnormal{d}\mathbf{X}\ |W(\mathbf{X})| \quad ,
\end{eqnarray}
and the exact fermionic expectation value of interest can then be computed as
\begin{align}
\langle O \rangle = \frac{\langle OS\rangle^\prime}{\langle S \rangle^\prime}\quad , 
\label{eq:average}
\end{align}
with averages being carried out over the modified distribution $W'(\mathbf{X}) = |W(\mathbf{X})|$ and $S=W(\mathbf{X})/|W(\mathbf{X})|=\textnormal{sgn}^\textnormal{f}(\sigma_\uparrow,\sigma_\downarrow)$ denoting the sign.
In practice, both the enumerator and denominator in Eq.~(\ref{eq:average}) vanish simultaneously with increasing system size and decreasing temperature. This leads to an exponentially increasing statistical uncertainty (error bar), which is nothing else than the notorious fermion sign problem~\cite{loh,troyer}. In fact, the FSP constitutes the paramount obstacle in our simulations and prevents PIMC simulations of the UEG for lower temperatures and higher densities, i.e., in the regime where quantum degeneracy effects dominate~\cite{dornheim_pop,review}.

\begin{figure}
\includegraphics[width=0.4147\textwidth]{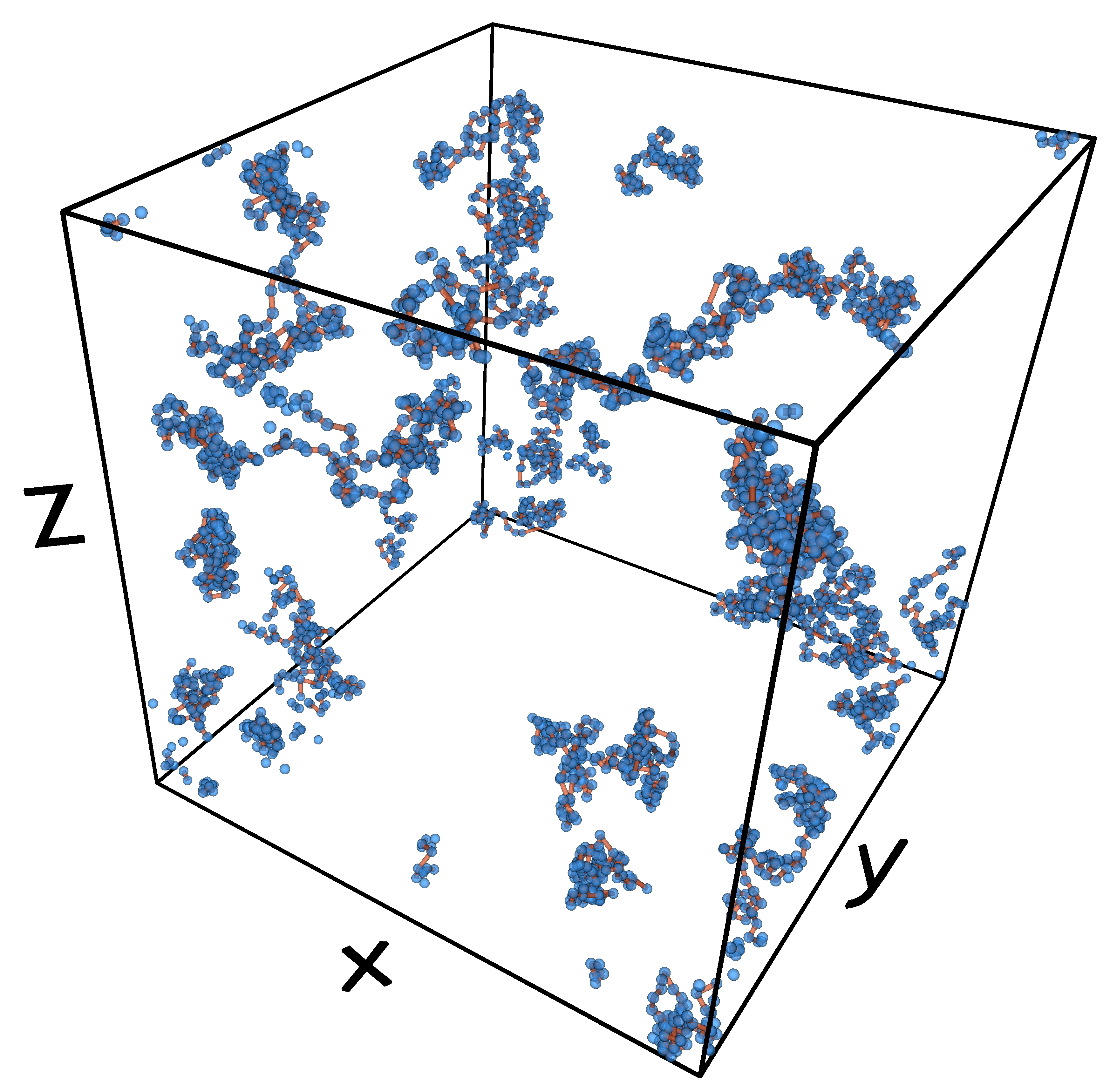}
\caption{\label{fig:PIMC_snapshot}
Snapshot of a PIMC simulation of the UEG at $r_s=1$ and $\theta=1$ with $N=17$ and $P=100$.
}
\end{figure}  
In Fig.~\ref{fig:PIMC_snapshot} we show a snapshot from a PIMC simulation of $N=17$ electrons at $r_s=1$ and $\theta=1$ with $P=100$ imaginary-time slices. Note that we use an adaption of the worm algorithm by Boninsegni \textit{et al.}~\cite{boninsegni1,boninsegni2} throughout this work. Evidently, there are many permutations of particle coordinates present (see also the recent discussion of permutation properties in Ref.~\cite{dornheim_permutation}) at these warm dense matter conditions and the sign of $W(\mathbf{X})$ within a simulation frequently changes. The resulting cancellation of positive and negative terms drastically increases the statistical uncertainty, which often cannot be compensated for by increasing the computation time, see Refs.~\cite{review, dornheim_pop} for a more extensive discussion.

For completeness, we mention that the more advanced permutation blocking PIMC~\cite{dornheim,dornheim2} and configuration PIMC~\cite{groth,dornheim3}
methods are capable to provide accurate results for static properties of the UEG when PIMC breaks down due to the sign problem. Unfortunately, however, the imaginary-time density-correlation function $F(\mathbf{q},\tau)$, which is of central importance in this work, can at present not be computed from these new techniques, so that PIMC remains the method of choice here.


\subsection{Linear response theory\label{sec:LRT}}

Let us next consider the effect of a small, time-dependent external perturbation, $\hat H_A(t)$, 
\begin{eqnarray}\label{eq:H_perturbed}
\hat H(t) = \hat H_\textnormal{UEG} + \hat H_A(t) \quad ,
\end{eqnarray}
with $\hat H_\textnormal{UEG}$ being the static UEG Hamiltonian. Note that we employ the standard Ewald summation as described in, e.g., Ref.~\cite{fraser}.
In particular, we consider a sinusoidal charge density of the form 
\begin{eqnarray}\label{eq:HA}
\hat H_A(t) = 2 A \sum_{i=1}^N \textnormal{cos}(\mathbf{r}_i\cdot \mathbf{q} - \Omega t) \quad ,
\end{eqnarray}
with the wave vector $\mathbf{q}$, frequency $\Omega$, and perturbation amplitude $A$.
Within linear response theory~\cite{quantum_theory}, which is accurate for sufficiently small perturbations, the response of the UEG to Eq.~(\ref{eq:HA}) is fully described by the density response function
\begin{eqnarray}\label{eq:chi_tilde}
\tilde \chi(\mathbf{q}, \overline{t}) = - \frac{i}{\hbar} \braket{[\rho(\mathbf{q},t),\rho(-\mathbf{q},t')]} \quad ,
\end{eqnarray}
where the expectation value $\braket{\dots}$ has to be carried out with respect to the unperturbed Hamiltonian, i.e., $\hat H_\textnormal{UEG}$. Therefore, Eq.~(\ref{eq:chi_tilde}) only depends on the relative difference between the time arguments, $\overline{t}=t-t'$, and on the modulus of the wave number $q=|\mathbf{q}|$.
We note that it is often more convenient to work in frequency space, and a straightforward Fourier transform gives
\begin{eqnarray}
\chi(\mathbf{q},\omega) = \lim_{\eta\to 0} \int_{-\infty}^\infty \textnormal{d}\overline{t}\ e^{(i\omega-\eta)\overline{t}}\tilde\chi(\mathbf{q},\overline{t}) \quad .
\end{eqnarray}
For completeness, we explicitly define the static density-response function, 
\begin{eqnarray}\label{eq:chi_static}
\chi(\mathbf{q}) = \lim_{\omega\to0} \chi(\mathbf{q},\omega) \quad ,
\end{eqnarray}
which describes the response of the UEG to a constant (i.e., time independent) external charge.

\subsection{Dynamic structure factor and imaginary-time correlation functions\label{sec:ITCF}}

A central quantity in modern WDM research (and many other fields) is the so-called dynamic structure factor,
\begin{eqnarray}\label{eq:dyn_def}
S(\mathbf{q},\omega) = \frac{1}{2\pi N} \int_{-\infty}^\infty \textnormal{d}t\ \braket{\rho(\mathbf{q},t)\rho(-\mathbf{q},0)} e^{i\omega t} \quad ,
\end{eqnarray}
which is directly accessible in XRTS experiments~\cite{siegfried_review} and is of paramount importance for diagnostics like, e.g., the determination of the electronic temperature within a sample~\cite{dominik}. We note that Eq.~(\ref{eq:dyn_def}) obeys the detailed balance condition
\begin{eqnarray}
S(\mathbf{q},\omega) =  S(\mathbf{q},-\omega) e^{-\beta\omega} \quad ,
\end{eqnarray}
so that the consideration of the positive frequency range is sufficient. In addition, $S(\mathbf{q},\omega)$ is directly connected to the imaginary part of the dynamic density-response function from the previous section by the fluctuation--dissipation theorem
\begin{eqnarray}\label{eq:FDT}
S(\mathbf{q},\omega) = - \frac{ \textnormal{Im}\chi(\mathbf{q},\omega)  }{ \pi n (1-e^{-\beta\omega})} \quad .
\end{eqnarray}

Evidently, $S(\mathbf{q},\omega)$ is nothing else than the Fourier transform of the intermediate scattering function
\begin{eqnarray}
F(\mathbf{q},t) = \frac{1}{N} \braket{\rho(\mathbf{q},t)\rho(-\mathbf{q},0)} \quad ,
\end{eqnarray}
and, in principle, requires an explicitly time-dependent theoretical description, which is notoriously difficult and almost unfeasible in the presence of correlation effects.

A neat alternative is given by the analytic continuation of the imaginary-time density correlation function, which is defined as
\begin{eqnarray}\label{eq:F}
F(\mathbf{q},\tau) = \frac{1}{N} \braket{\rho(\mathbf{q},\tau)\rho(-\mathbf{q},0)} \quad .
\end{eqnarray}
Although Eq.~(\ref{eq:F}) is formally obtained by a propagation in imaginary-time by $\tau = -i\hbar\beta$, the pre-factors are usually dropped such that $\tau\in[0,\beta]$; this convention is applied throughout the remainder of this paper. 
The crucial point in the context of the present work is that $F(\mathbf{q},\tau)$ is directly accessible within our PIMC simulations in thermodynamic equilibrium~\cite{berne1,berne2} by measuring the correlation between the Fourier components of the density operator on different imaginary-time slices, see Fig.~\ref{fig:PIMC} for a graphical depiction.
The connection of $F$ to the dynamic structure factor is then given by a Laplace transform
\begin{eqnarray}\label{eq:FS}
F(\mathbf{q},\tau) = \int_{-\infty}^\infty \textnormal{d}\omega\ S(\mathbf{q},\omega) e^{-\tau\omega} \quad .
\end{eqnarray}
Therefore, the task at hand is to numerically solve Eq.~(\ref{eq:FS}) by performing an inverse Laplace transform, which is a well-known but ill-posed problem~\cite{jarrell}.

\subsection{Reconstruction of the dynamic structure factor\label{sec:reconstruction_results}}

The central obstacle regarding the reconstruction of the dynamic structure factor from an imaginary-time correlation function is the statistical uncertainty in the corresponding QMC data. Therefore, there are potentially infinitely many valid trial solutions $S_\textnormal{trial}(\mathbf{q},\omega)$, which, when being inserted into Eq.~(\ref{eq:FS}), perfectly reproduce the Monte Carlo data within the given error bars. Typically, these $S_\textnormal{trial}(\mathbf{q},\omega)$ are very noisy (\textit{sawtooth instability}) and often plainly unphysical. Hence, additional constraints on the trial solutions are indispensable to obtain reliable structure factors on the basis of our PIMC data.

A commonly used type of information are frequency moments of the form
\begin{eqnarray}
\braket{\omega^k} = \int_{-\infty}^\infty \textnormal{d}\omega\ S(\mathbf{q},\omega)\ \omega^k \quad ,
\end{eqnarray}
where, in the case of the UEG, four different cases are known:
\begin{enumerate}
    \item The inverse moment is determined by the static density response function~\cite{iwamoto2,pines_woo,kugler2}
\begin{eqnarray}
\braket{\omega^{-1}} = - \frac{\chi(\mathbf{q})}{2n} \quad ,
\end{eqnarray}
see also Eq.~(\ref{eq:chi_static}). Conveniently, $\chi(\mathbf{q})$ is straightforwardly obtained from $F(\mathbf{q},\tau)$ via the imaginary-time analogue of the fluctuation--dissipation theorem, which states that~\cite{bowen,alex_prok}
\begin{eqnarray}\label{eq:static_chi}
\chi(\mathbf{q}) = -n\int_0^\beta \textnormal{d}\tau\ F(\mathbf{q},\tau) \quad .
\end{eqnarray}

\item The normalization, or zero-moment, is given by the static structure factor
\begin{eqnarray}
\braket{\omega^0}  = \int_{-\infty}^\infty \textnormal{d}\omega\ S(\mathbf{q},\omega) =: S(\mathbf{q}) \quad .
\end{eqnarray}
Note that $S(\mathbf{q})$ is defined as the integral over $S(\mathbf{q},\omega)$ and not as the static limit as in the case of $\chi(\mathbf{q})$ and $\chi(\mathbf{q},\omega)$.

\item The first moment is known analytically from the f sum-rule~\cite{quantum_theory},
    \begin{eqnarray}
\braket{\omega^1} = \frac{q^2}{2} \quad .
\end{eqnarray}

\item The third moment was first reported in Refs.~\cite{puff1,puff2} and reads~\cite{iwamoto2,iwamoto,quantum_theory}
\begin{eqnarray}
\braket{\omega^3} = \frac{q^2}{2} \Bigg(
&\Big(&\frac{q^2}{2}\Big)^2 + q^2 n v_q + 2q^2K\\ \nonumber &+& \omega_p^2\big(1-I(q)\big)
\Bigg) \quad ,
\end{eqnarray} with $K$ being the mean kinetic energy of the correlated system, and
where the potential contribution~\cite{iwamoto,quantum_theory} can be expressed in spherical coordinates as a one-dimensional integral,
 \begin{eqnarray}\label{eq:I}
 I(q) &=&  \frac{1}{8\pi^2n} \int_0^\infty \textnormal{d}k\ k^2 \big(1-S(k)\big) \\ \nonumber
 &\times& \Bigg( 
 \frac{5}{3} - \frac{k^2}{q^2} + \frac{\big(k^2-q^2\big)^2}{2kq^3} \textnormal{log}\Bigg| \frac{k+q}{k-q} \Bigg| 
 \Bigg) \quad ,
 \end{eqnarray}
which is evaluated numerically from spline-fits to $S(\mathbf{q})$, see also Refs.~\cite{dornheim_prl,dornheim_cpp} for an extensive discussion.

\end{enumerate}

While accurate knowledge of the $\braket{\omega^k}$ are known to significantly increase the quality of the reconstructed dynamic structure factors for some examples such as ultracold atoms~\cite{dynamic_alex1,dynamic_alex2}, they have been proven to be insufficient to determine sufficiently constrained $S_\textnormal{trial}(\mathbf{q},\omega)$ for the UEG in many cases.

\subsection{Stochastic sampling of the dynamic local field correction\label{sec:stochastic_sampling}}

To derive additional constraints on the reconstructed dynamic structure factors, we consider the fluctuation--dissipation theorem, Eq.~(\ref{eq:FDT}), and express $\chi(\mathbf{q},\omega)$ in terms of a dynamic local field correction~\cite{kugler1,quantum_theory,gross,dabrowski} $G(\mathbf{q},\omega)$
\begin{eqnarray}\label{eq:define_LFC}
\chi(\mathbf{q},\omega) = \frac{ \chi_0(\mathbf{q},\omega) }{ 1 - 4\pi/q^2\big[1-G(\mathbf{q},\omega)\big]\chi_0(\mathbf{q},\omega)} \quad .
\end{eqnarray}
The dynamic density-response function of the noninteracting system, $\chi_0(\mathbf{q},\omega)$, is readily known, and all exchange-correlation effects regarding the density response are contained in $G$. Therefore, setting $G=0$ corresponds to the random phase approximation, which describes the response on a mean-field level.
In a nutshell, the combination of Eqs.~(\ref{eq:FDT}) and (\ref{eq:define_LFC}) implies that we have re-cast the reconstruction problem from a quest for $S(\mathbf{q},\omega)$ into a quest for the DLFC. This is extremely advantageous, as many additional exact properties of $G$ are known:
\begin{enumerate}
    \item The Kramers-Kronig relations provide a connection between the real and imaginary parts of $G(\mathbf{q},\omega)$~\cite{kugler1}:
\begin{eqnarray}\label{eq:Kramers_Kronig_real}
\textnormal{Re}G(\mathbf{q},\omega) &=& \textnormal{Re}G(\mathbf{q},\infty) + \frac{1}{\pi} \int_{-\infty}^\infty \textnormal{d}\overline{\omega}\ \frac{\textnormal{Im}G(\mathbf{q},\overline{\omega})}{\overline{\omega}-\omega} \\
\textnormal{Im}G(\mathbf{q},\omega) &=& \\ \nonumber &-& \frac{1}{\pi} \int_{-\infty}^\infty \textnormal{d}\overline{\omega}\ 
\frac{\textnormal{Re} G(\mathbf{q}, \overline{\omega}) - \textnormal{Re}G(\mathbf{q},\infty)  }{\overline{\omega}-\omega} \quad .
\end{eqnarray}

\item $\textnormal{Re}G(\mathbf{q},\omega)$ and $\textnormal{Im}G(\mathbf{q},\omega)$ are even and odd functions with respect to $\omega$, respectively~\cite{gross}.
\item  $\textnormal{Im}G(\mathbf{q},\omega)$  vanishes in the limits of high and low frequency~\cite{gross}:
\begin{eqnarray}
\textnormal{Im}G(\mathbf{q},0) = \textnormal{Im}G(\mathbf{q},\infty) = 0 \quad .
\end{eqnarray}

\item The static limit of $\textnormal{Re}G(\mathbf{q},\omega)$ is defined by static density response function [see Eq.~(\ref{eq:static_chi})], since $\chi(\mathbf{q})$ is a real function~\cite{quantum_theory}:
\begin{eqnarray}\label{eq:G_static}
\textnormal{Re}G(\mathbf{q},0) = 1 - \frac{1}{v_q}\left( 
\frac{1}{\chi_0(\mathbf{q},0)} - \frac{1}{\chi(\mathbf{q})}
\right) \quad .
\end{eqnarray}

The high-frequency limit of $\textnormal{Re}G(\mathbf{q},\omega)$ can be computed from the static structure factor $S(\mathbf{q})$ and the exchange-correlation contribution to the kinetic energy, $K_\textnormal{xc}$,
 \begin{eqnarray}\label{eq:G_infty}
 \textnormal{Re}G(\mathbf{q},\infty) = I(q) - \frac{ 2 q^2 K_\textnormal{xc}}{\omega_p^2} \quad .
 \end{eqnarray}
The interaction contribution $I(q)$ has been defined in Eq.~(\ref{eq:I}), and $K_\textnormal{xc}$ can be computed from the exchange-correlation free energy $f_\textnormal{xc}$ via
 \begin{eqnarray}\label{eq:Kxc}
 K_\textnormal{xc} 
 &=& - f_\textnormal{xc}(r_s,\theta)
 -\theta \frac{\partial f_\textnormal{xc}(r_s,\theta)}{\partial\theta}\Bigg|_{r_s}\\ \nonumber
& & - r_s \frac{\partial f_\textnormal{xc}(r_s,\theta)}{\partial r_s}\Bigg|_\theta \quad .
 \end{eqnarray}
In practice, we are using the accurate recent parametrization of $f_\textnormal{xc}$ by Groth, Dornheim and co-workers (see Refs.~\cite{groth_prl,review}) to evaluate Eq.~(\ref{eq:Kxc}).

\end{enumerate}

The basic idea of our new reconstruction method is to stochastically sample trial solutions $G_\textnormal{trial}(\mathbf{q},\omega)$, which automatically fulfill all aforementioned exact properties.
These DLFCs are then used to compute the corresponding trial DSFs $S_\textnormal{trial}(\mathbf{q},\omega)$, which are subsequently compared to our PIMC data for $F(\mathbf{q},\tau)$ and the four frequency moments $\braket{\omega^k}$. 
To accomplish the first part of this task, we are introducing extended Pad\'e type parametrizations of the imaginary part of the DLFC of the form
\begin{eqnarray}\label{eq:parametrization}
\textnormal{Im}G(\mathbf{q},\omega) = \frac{ a_0\omega + a_1\omega^3 + a_2\omega^5 }{ \left( b_0 + b_1\omega^2 \right)^c } \quad ,
\end{eqnarray}
with $a_i$, $b_i$, and $c$ being the free parameters. Once the latter have been randomly generated, the real part of $G(\mathbf{q},\omega)$ is obtained numerically from the Kramers-Kronig relation, Eq.~(\ref{eq:Kramers_Kronig_real}), and imposing the exact static limit of $\textnormal{Re}G(\mathbf{q},\omega)$ uniquely determines one free parameter in terms of the others, 
\begin{eqnarray}\label{eq:limit_integral}
\textnormal{Re}G(\mathbf{q},0) &\stackrel{!}{=}& \textnormal{Re}G(\mathbf{q},\infty)\\ \nonumber & & + \frac{1}{\pi}\int_{-\infty}^\infty \textnormal{d}{\omega}\ \frac{ a_0 + a_1\omega^2 + a_2\omega^4 }{ \left( b_0 + b_1\omega^2 \right)^c  } \quad .
\end{eqnarray}
In practice, the integral in Eq.~(\ref{eq:limit_integral}) was solved analytically using SymPy~\cite{sympy} to obtain the parameter $a_1$.
The remaining five free parameters are randomly chosen from the following empirically found intervals: $a_0,a_2,b_1\in[10^{-5},10^{5}]$, $a_1\in[10^{-3},10^{5}]$, and $c\in[\frac{5}{2},25]$.

The final result for the dynamic structure factor is then computed as the average over $M$ trial spectra $S_{\textnormal{trial},i}(\mathbf{q},\omega)$ that reproduce the $\braket{\omega^k}$ and $F(\mathbf{q},\tau)$ for all $P$ $\tau$-points within the Monte Carlo error bars,
\begin{eqnarray}\label{eq:final_average}
S_\textnormal{final}(\mathbf{q},\omega) = \frac{1}{M} \sum_{i=1}^M S_{\textnormal{trial},i}(\mathbf{q},\omega) \quad ,
\end{eqnarray}
where it typically holds $M=\mathcal{O}(1000)$.
In addition, this allows for a straightforward estimation of the remaining uncertainty of the reconstruction by obtaining the corresponding variance
\begin{eqnarray}\label{eq:delta_S}
\Delta S(\mathbf{q},\omega) = \left( 
\frac{1}{M}\sum_{i=1}^M\left[ 
S_{\textnormal{trial},i}(\mathbf{q},\omega) \right.\right. \\  - \left.\left. S_\textnormal{final}(\mathbf{q},\omega)
\right]^2
\right)^{1/2} \nonumber \quad .
\end{eqnarray}
The practical application of this new reconstruction method is extensively demonstrated in Sec.~\ref{sec:stochastic_sampling_results}.

\section{Results\label{sec:results}}

\subsection{Imaginary-time density--density correlation function\label{sec:F_results}}

\begin{figure}\vspace*{-0.752cm}\hspace*{-0.3cm}\includegraphics[width=0.516\textwidth]{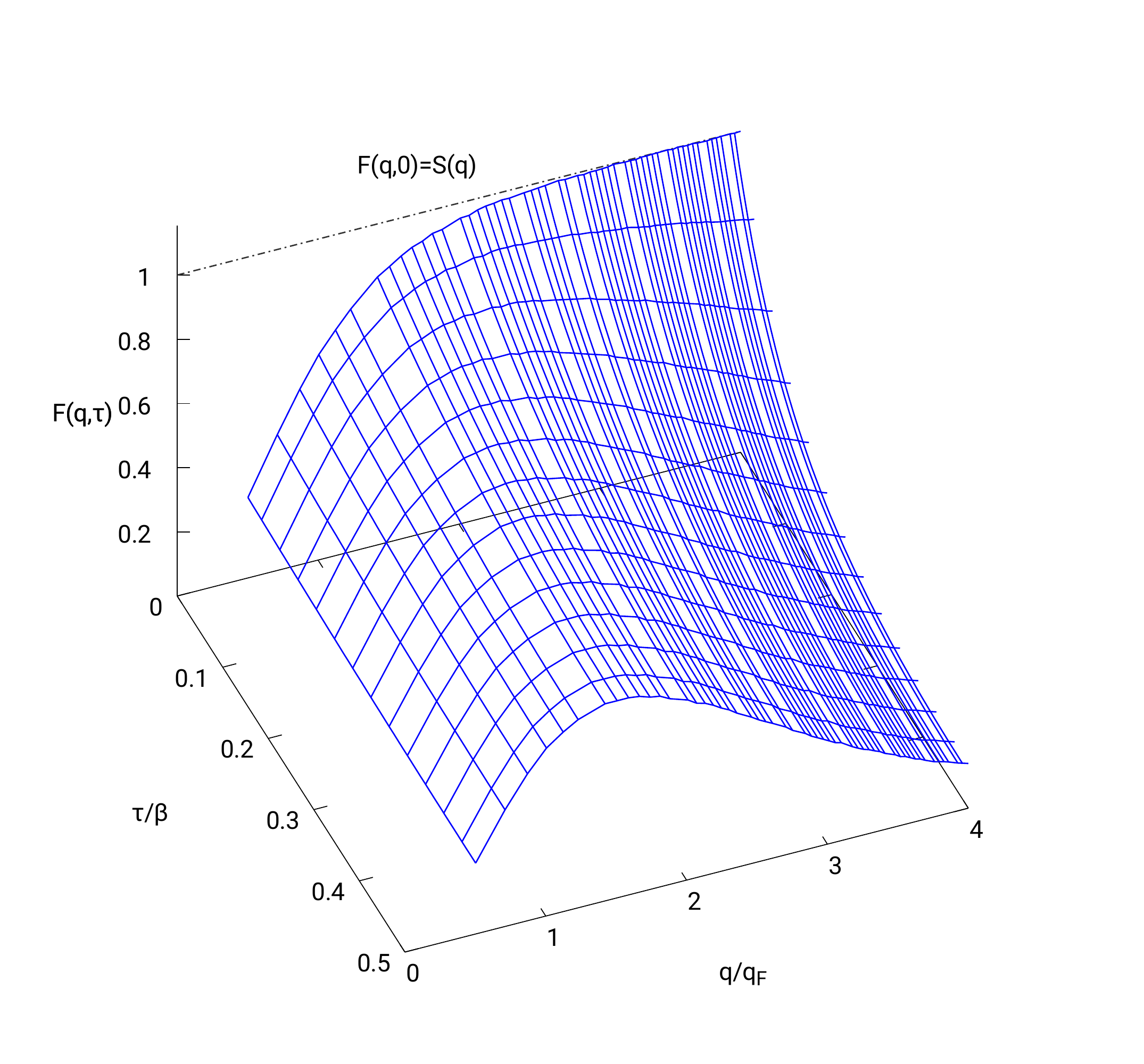}\\ \vspace*{-1.22cm}
\includegraphics[width=0.516\textwidth]{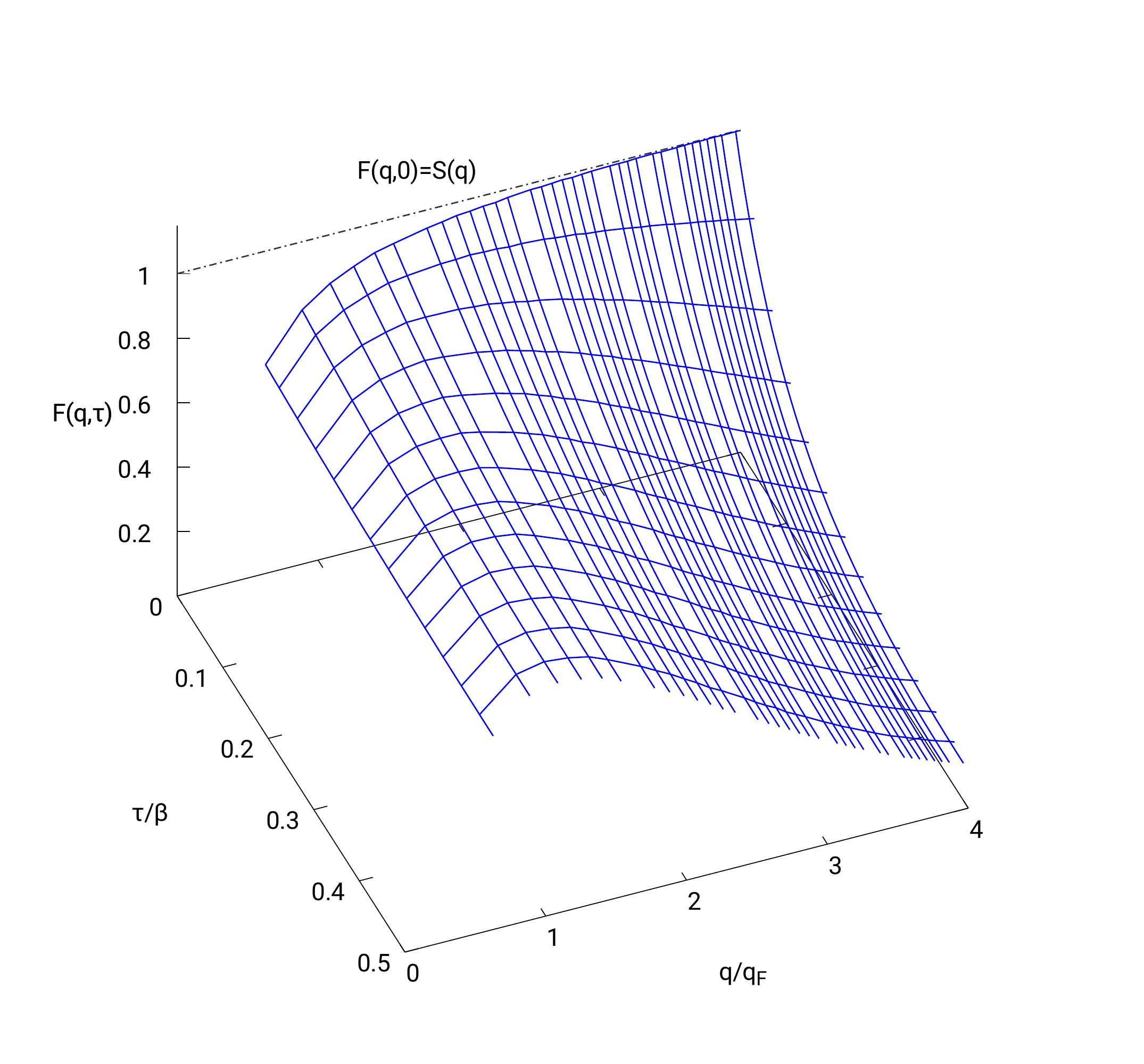}\vspace*{-0.8cm}\caption{\label{fig:FPLOT_theta2}Imaginary-time density--density correlation function $F(q,\tau)$ for $\theta=2$ and $r_s=4$, $N=66$ (top) and $r_s=1$, $N=34$ (bottom). $F(q,\tau)$ is symmetric with respect to $\tau=\beta/2$ and becomes equivalent to the static structure factor $S(q)$ for $\tau=0$.}\end{figure}

Let us start the discussion of our simulation results by considering the central quantity that can be directly obtained from PIMC simulations of the UEG, namely the imaginary-time density--density correlation function $F(\mathbf{q},\tau)$.
\begin{figure}
\includegraphics[width=0.4147\textwidth]{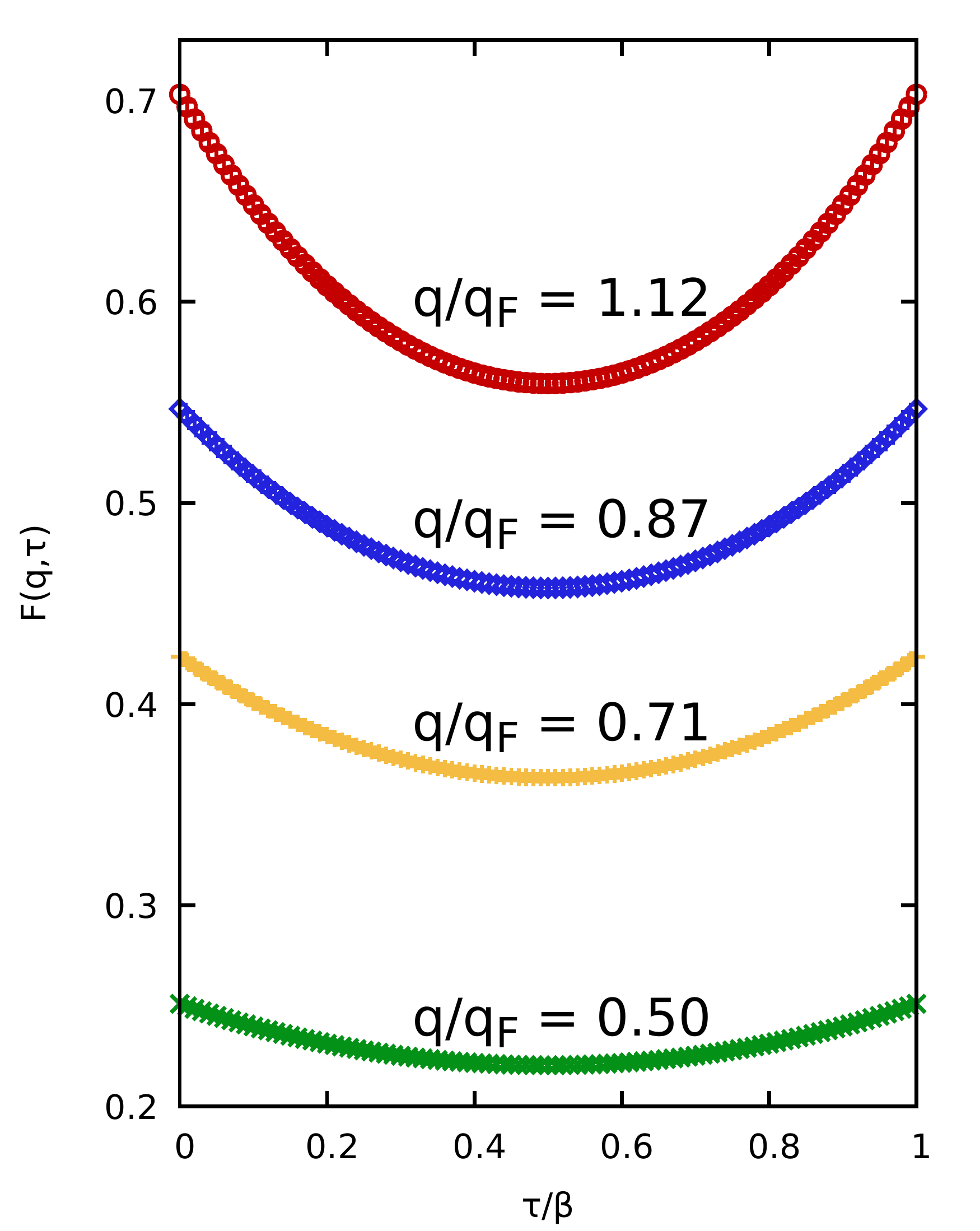}
\caption{\label{fig:ITCF_rs4_theta2}
\textit{Ab initio} PIMC results for the imaginary-time density--density correlation function $F(q,\tau)$ for fixed wave numbers at $r_s=4$, $\theta=2$, and $N=66$.
}
\end{figure}  
In Fig.~\ref{fig:ITCF_rs4_theta2}, we show results for the $\tau$-dependence of this quantity for $N=66$ unpolarized electrons at $r_s=4$ and $\theta=2$ for four different wave numbers $q=|\mathbf{q}|$ in the vicinity of the Fermi wave number $q_\textnormal{F}$.
First and foremost, we note that $F$ is always symmetric with respect to $\tau=\beta/2$, i.e., $F(\mathbf{q},\tau) = F(\mathbf{q},\beta-\tau)$ (for $\tau \leq \beta/2$). In addition, the magnitude of $F$ increases monotonically with increasing $q$ in the depicted regime of wave vectors. This can be understood by recalling its relation to the static structure factor, $F(\mathbf{q},\tau=0) = S(\mathbf{q})$, with $S(\mathbf{q})$ following a parabola for small $q$ and being monotonically increasing up to around thrice the Fermi wave number, see Fig.~\ref{fig:FPLOT_theta2} for a graphical depiction and Refs.~\cite{dornheim_prl,dornheim_cpp,review} for an extensive discussion. Lastly, we mention that the slope of $F$ becomes increasingly steep for larger $q$; a trend that persists for even larger wave numbers as can be seen in Fig.~\ref{fig:ITCF_rs4_theta2_large} where we show PIMC results for up to $q/q_\textnormal{F}\approx6$.
\begin{figure}
\includegraphics[width=0.4147\textwidth]{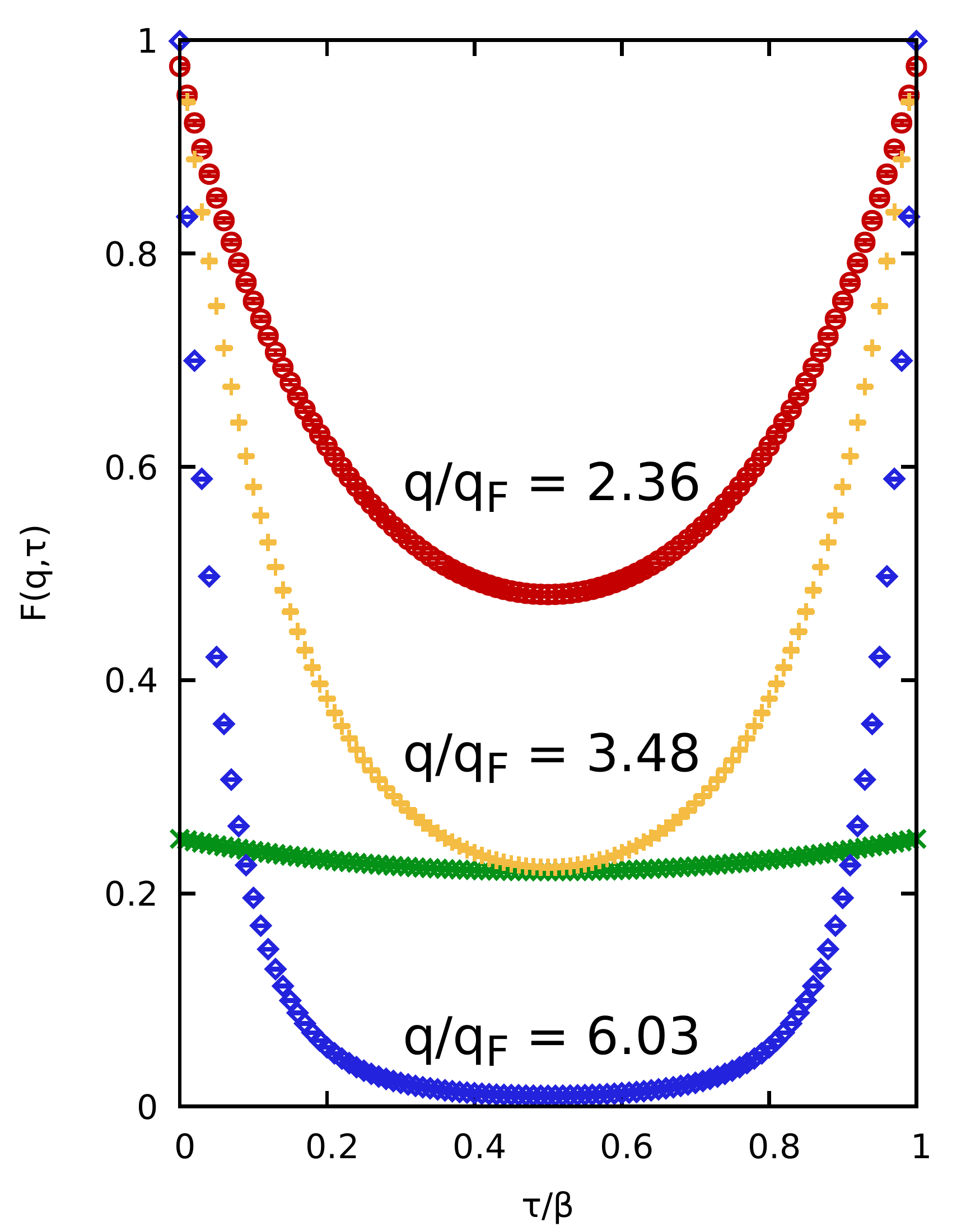}
\caption{\label{fig:ITCF_rs4_theta2_large}
Same as Fig.~\ref{fig:ITCF_rs4_theta2}, but for larger wave numbers. The green curve is shown as a reference and corresponds to $q/  q_\textnormal{F}=0.50$, i.e., to the green curve in Fig.~\ref{fig:ITCF_rs4_theta2}.
}
\end{figure}  
In particular, $F(\mathbf{q},\tau)$ almost decays to zero around $\tau=\beta/2$ for large $q$.

\begin{figure}
\includegraphics[width=0.400147\textwidth]{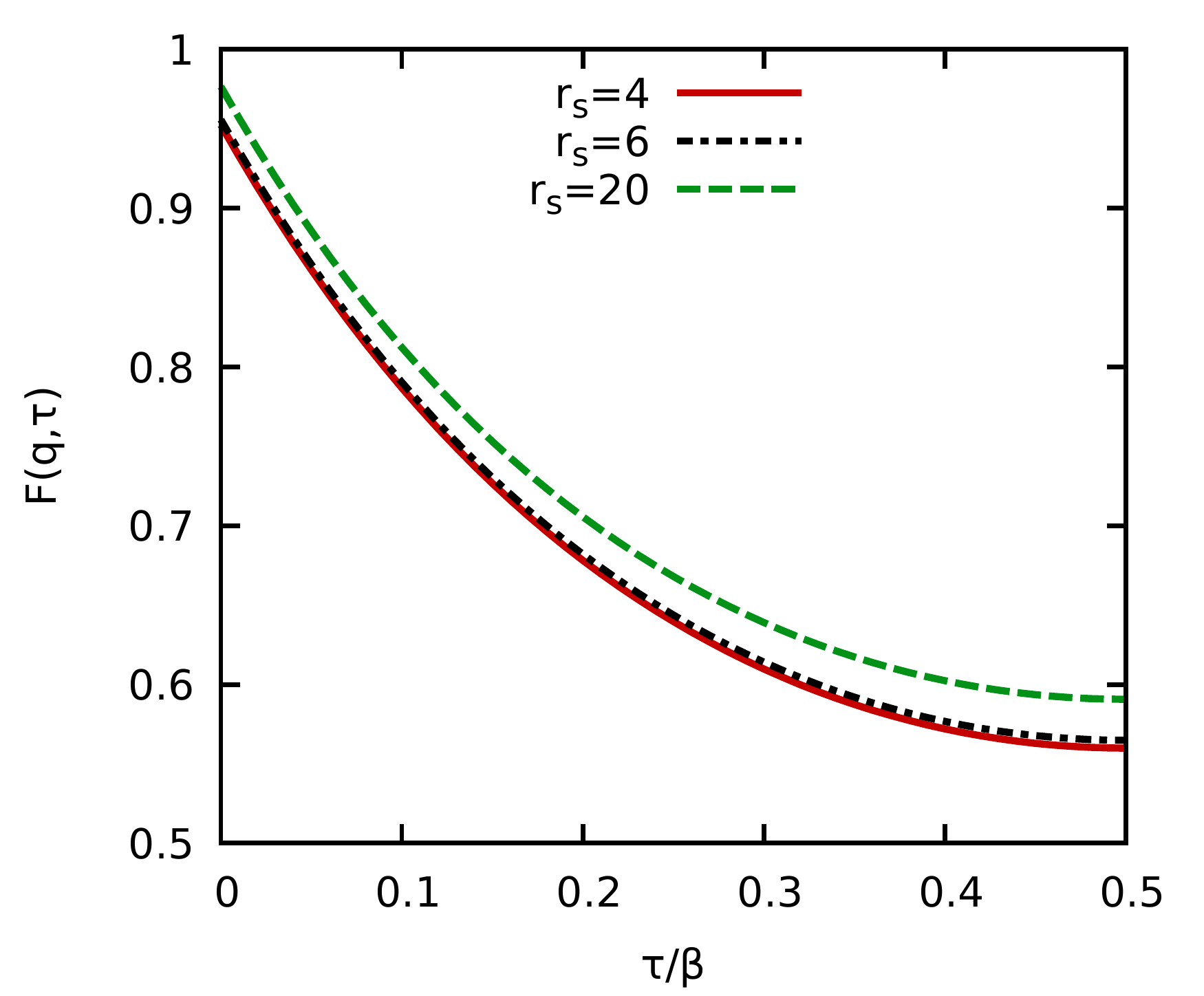}
\caption{\label{fig:ITCF_theta2_rs}
Imaginary-time density--density correlation function $F(q=2q_\textnormal{F},\tau)$ at $\theta=2$ and $N=66$ for $r_s=4$ (solid red), $r_s=6$ (dash-dotted black), and $r_s=20$ (dashed green).
}
\end{figure}

Let us next study the impact of Coulomb coupling effects by considering different values of the density parameter $r_s$.
In Fig.~\ref{fig:ITCF_theta2_rs}, we compare our results for $F(\mathbf{q},\tau)$ again for $N=66$ unpolarized electrons at $\theta=2$ for $r_s=4$ (solid red), $r_s=6$ (dash-dotted black), and $r_s=20$ (dashed green).
The wave number has been chosen as $q/q_\textnormal{F}\approx 2$ corresponding to the physically most interesting regime where the dispersion relation of $S(\mathbf{q},\omega)$ is actually negative for strong coupling, like in the case of $r_s=20$, see Fig.~\ref{fig:breit}.
Indeed, while the curves for $r_s=4$ and $r_s=6$ are nearly identical, the $r_s=20$ results significantly differ both in magnitude and slope, cf.~the corresponding dynamic structure factors $S(\mathbf{q},\omega)$ shown in Fig.~\ref{fig:breit}.

\begin{figure}
\includegraphics[width=0.400147\textwidth]{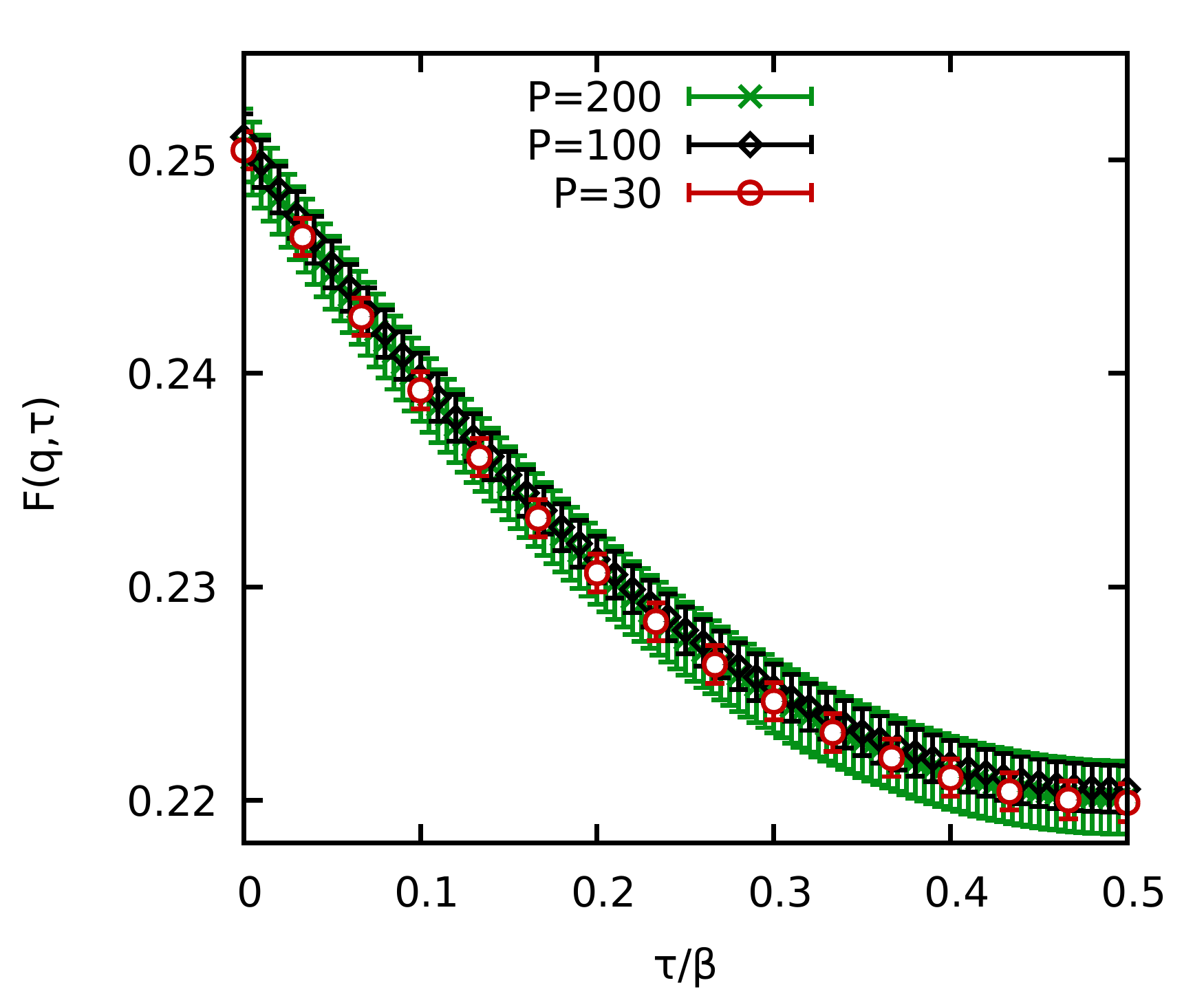}
\caption{\label{fig:ITCF_theta2_rs4_P}
Imaginary-time density--density correlation function $F(q=0.5q_\textnormal{F},\tau)$ at $\theta=2$, $r_s=4$, and $N=66$ for different numbers of imaginary-time propagators $P$.
}
\end{figure}  
For completeness, let us also consider the dependence of the imaginary-time density correlation function on the number of propagators $P$. This is investigated in Fig.~\ref{fig:ITCF_theta2_rs4_P}, where we show results for $F$ for $N=66$ electrons at $r_s=4$ and $\theta=2$ for $q/q_\textnormal{F}\approx0.5$. The different symbols correspond to PIMC data for $P=30$ (red circles), $P=100$ (black diamonds), and $P=200$ (green crosses, in the background). Evidently, the four different data sets cannot be distinguished within the statistical uncertainty and the only difference is given by the different $\tau$-grid.
The situation somewhat changes towards lower temperatures and larger $q$, and the convergence with $P$ was carefully checked for all considered cases throughout this work.

Let us conclude our consideration of $F(\mathbf{q},\tau)$ with a depiction of this quantity over the entire relevant $\tau$-$q$-plane.
This is shown for $N=66$, $\theta=2$, and $r_s=4$ in the top panel of Fig.~\ref{fig:FPLOT_theta2}. Since a physically meaningful interpretation of this quantity is rather difficult, we restrict ourselves to mentioning that $F$ converges to the static structure factor for $\tau=0$ and exhibits a distinct structure around $q=2q_\textnormal{F}$. The bottom panel of the same figure corresponds to $N=34$, $\theta=2$, and $r_s=1$. At this higher density, correlation effects are less important and the structure in $F(\mathbf{q},\tau)$ becomes less pronounced.

\begin{figure}\vspace*{-0.752cm}\hspace*{-0.3cm}\includegraphics[width=0.516\textwidth]{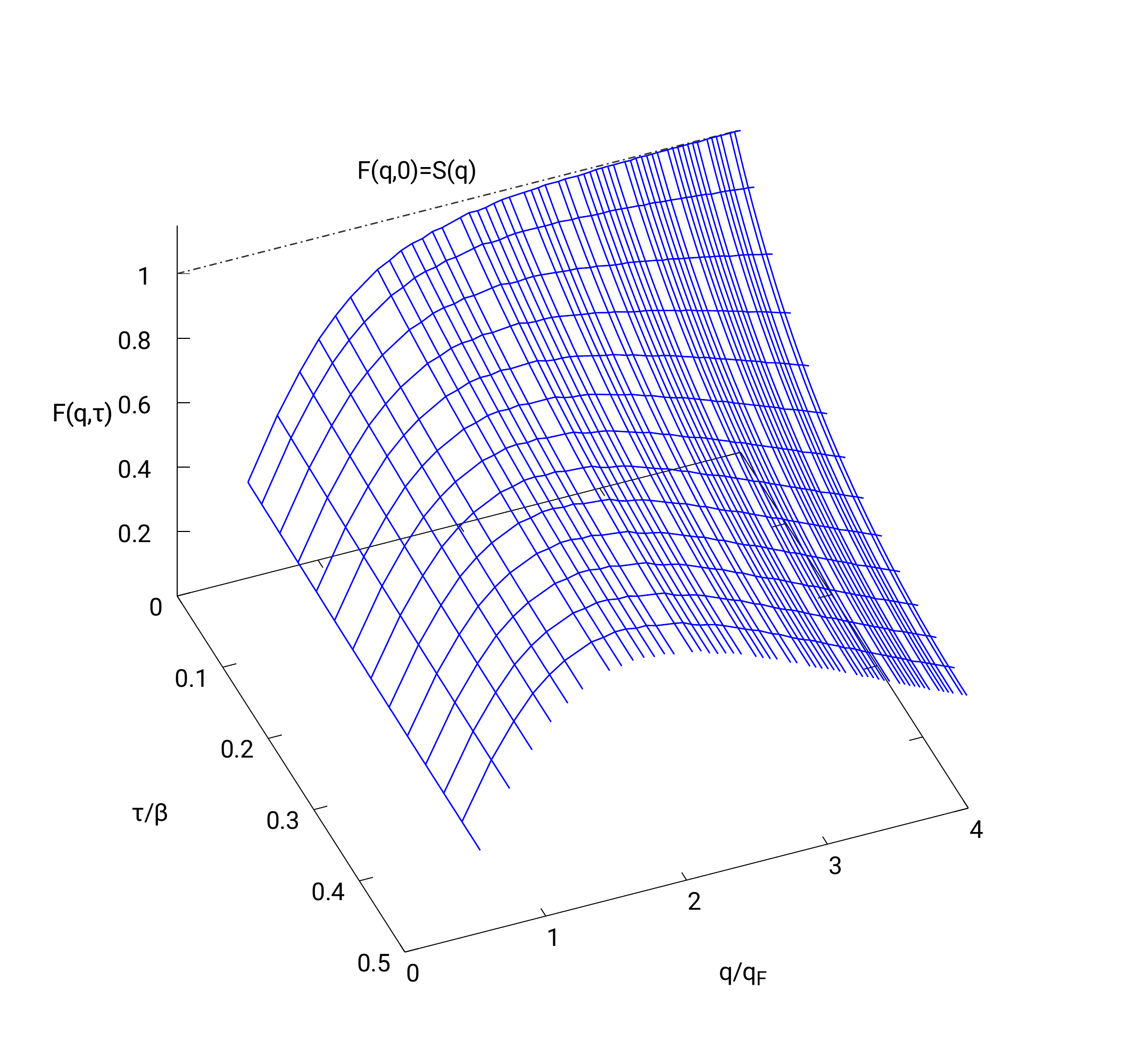}\\
\vspace*{-1.22cm}\hspace*{-0.3cm}\includegraphics[width=0.516\textwidth]{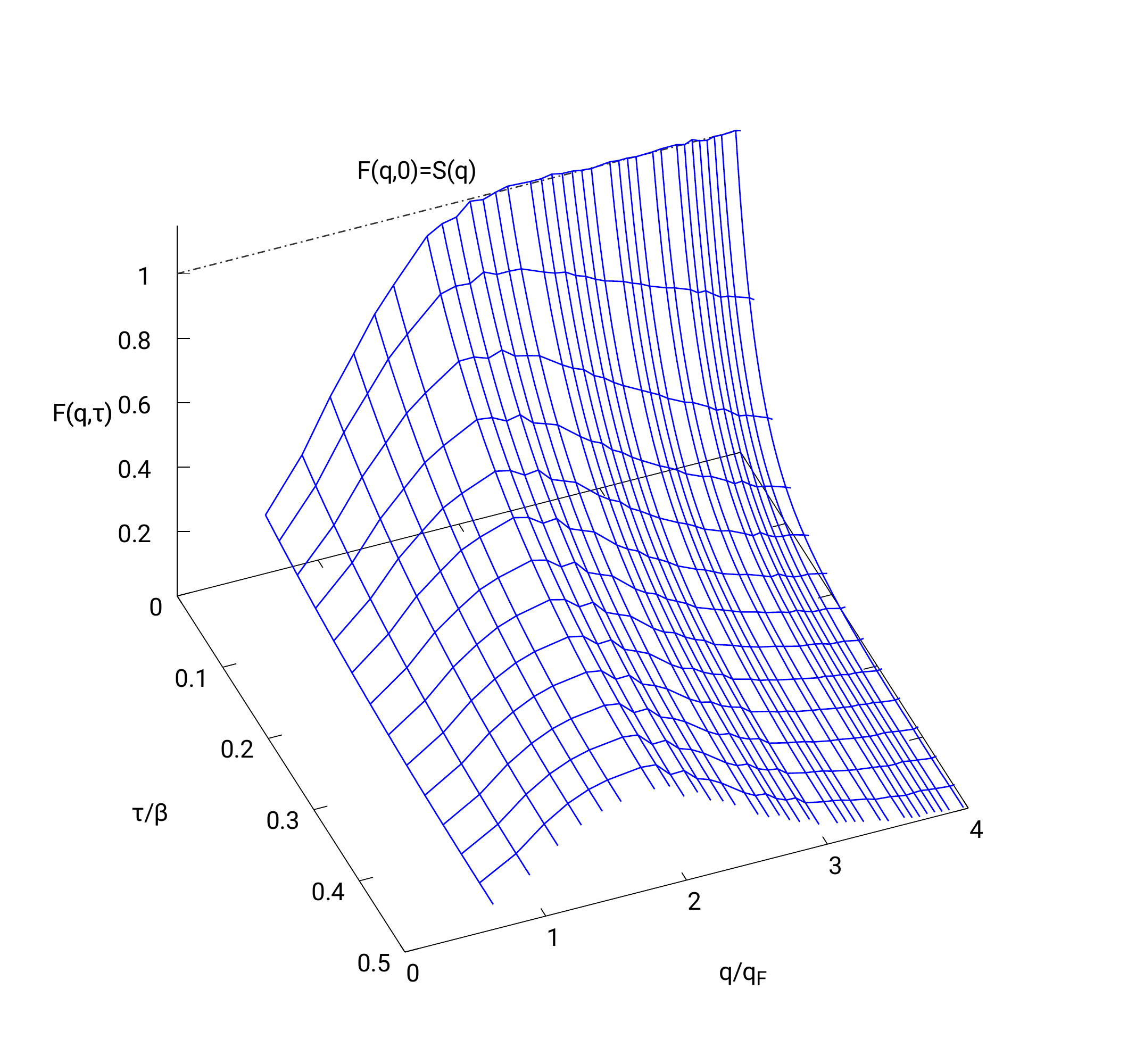}\vspace*{-0.8cm}\caption{\label{fig:FPLOT_rs6}Imaginary-time density--density correlation function $F(q,\tau)$ for $r_s=6$ and $\theta=4$, $N=66$ (top) and $\theta=0.75$, $N=34$ (bottom).}\end{figure}  
Lastly, we show PIMC results for $F$ for $r_s=6$ and $\theta=4$ (Fig.~\ref{fig:FPLOT_rs6}, top) and $\theta=0.75$ (bottom). Again, we observe that the larger impact of correlation effects leads to a significantly more structured imaginary-time correlation function at the lower temperature. The somewhat noisy progression in the case of $\theta=0.75$ around $q=2q_\textnormal{F}$ is a result of the increased statistical uncertainty in our PIMC data due to the fermion sign problem with an average sign of $S\approx0.023$.

\subsection{Static density-response function and local field corrections\label{sec:static_response}}

Another important quantity that is of considerable interest in its own right is the static density--density response function $\chi(\mathbf{q})$, which is obtained from $F(\mathbf{q},\tau)$ via a simple one-dimensional integration along the $\tau$-axis, see Eq.~(\ref{eq:static_chi}).
\begin{figure}
\includegraphics[width=0.4147\textwidth]{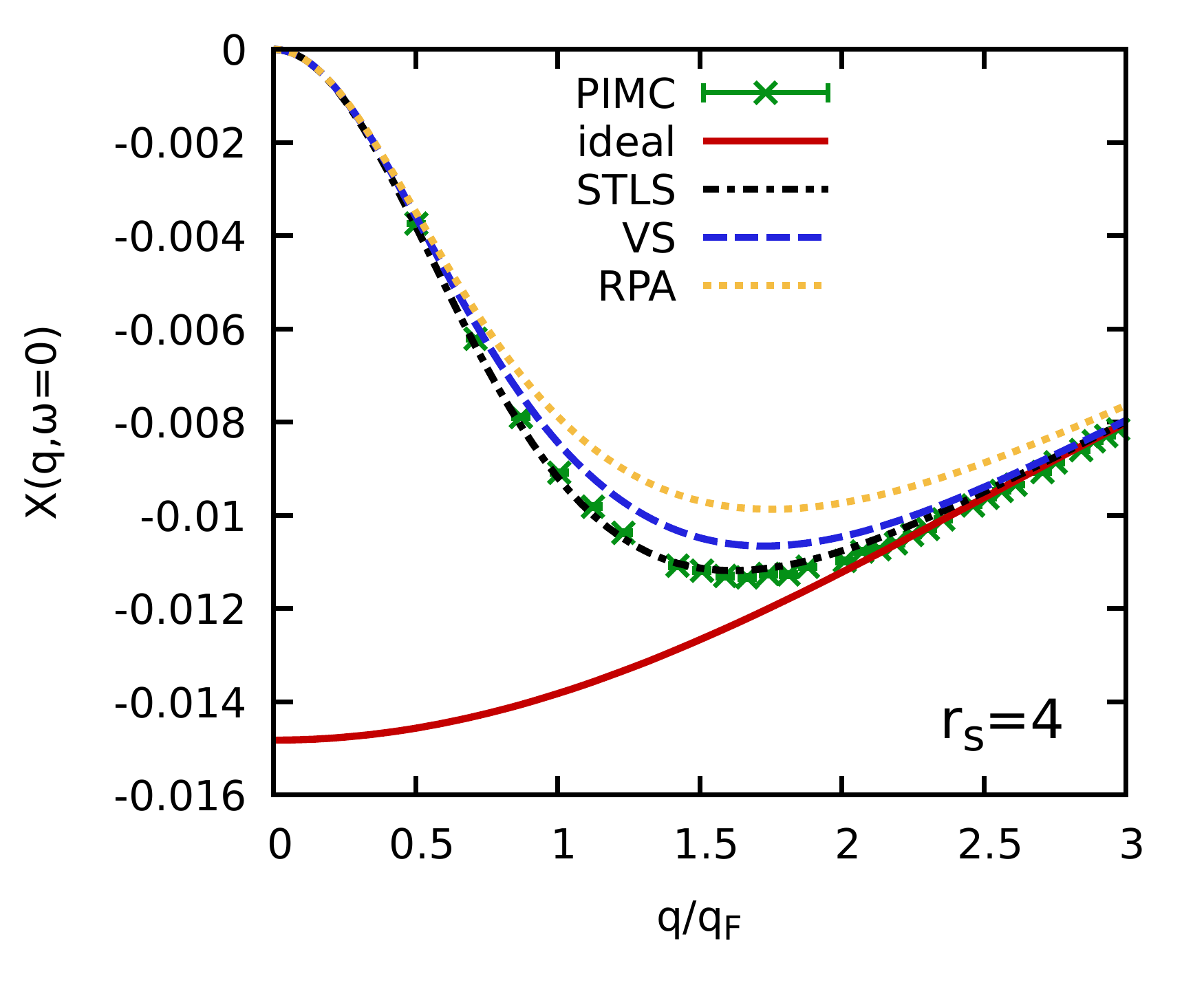}\\ \vspace*{-0.3cm}
\includegraphics[width=0.4147\textwidth]{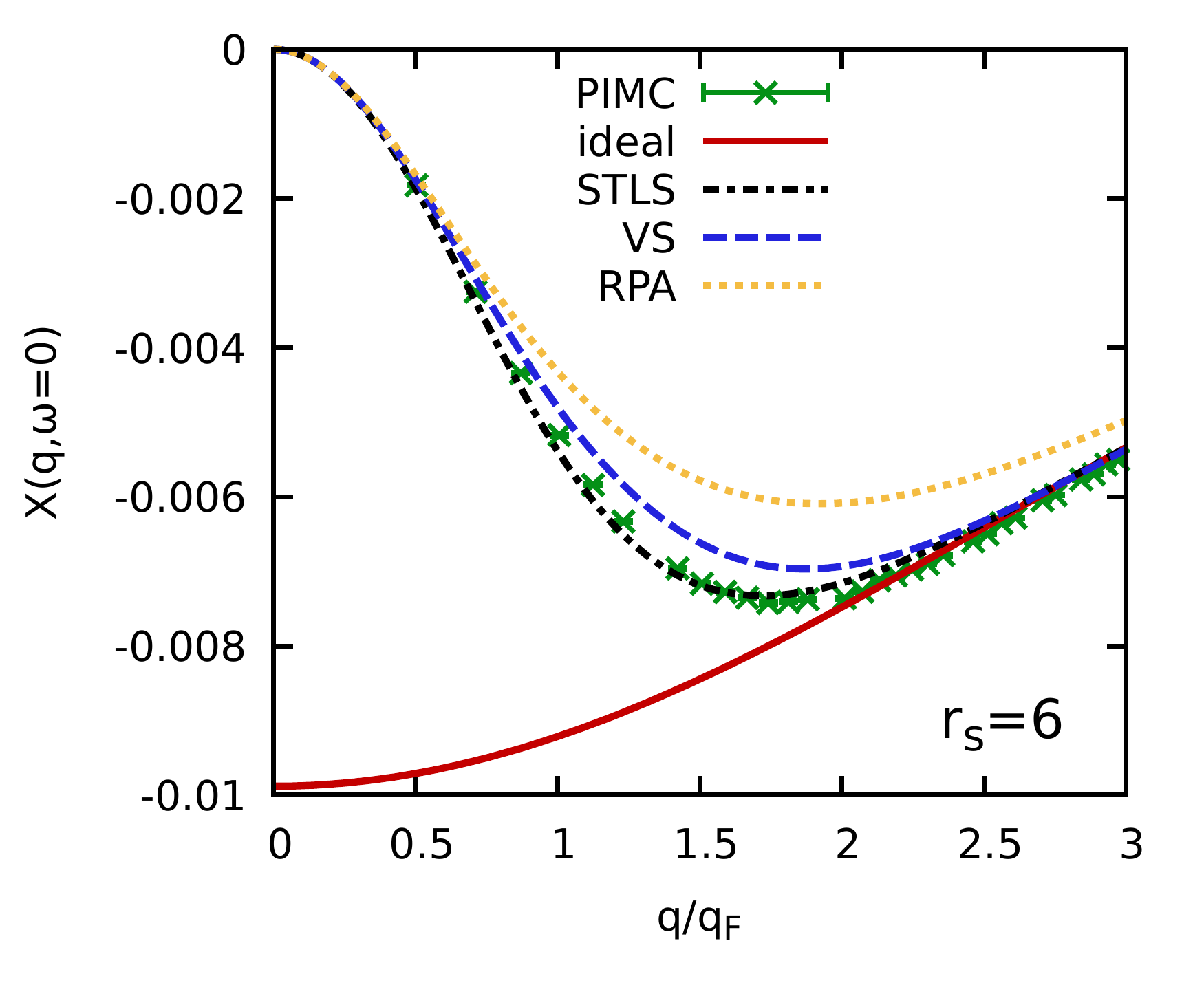}\\ \vspace*{-0.3cm}
\includegraphics[width=0.4147\textwidth]{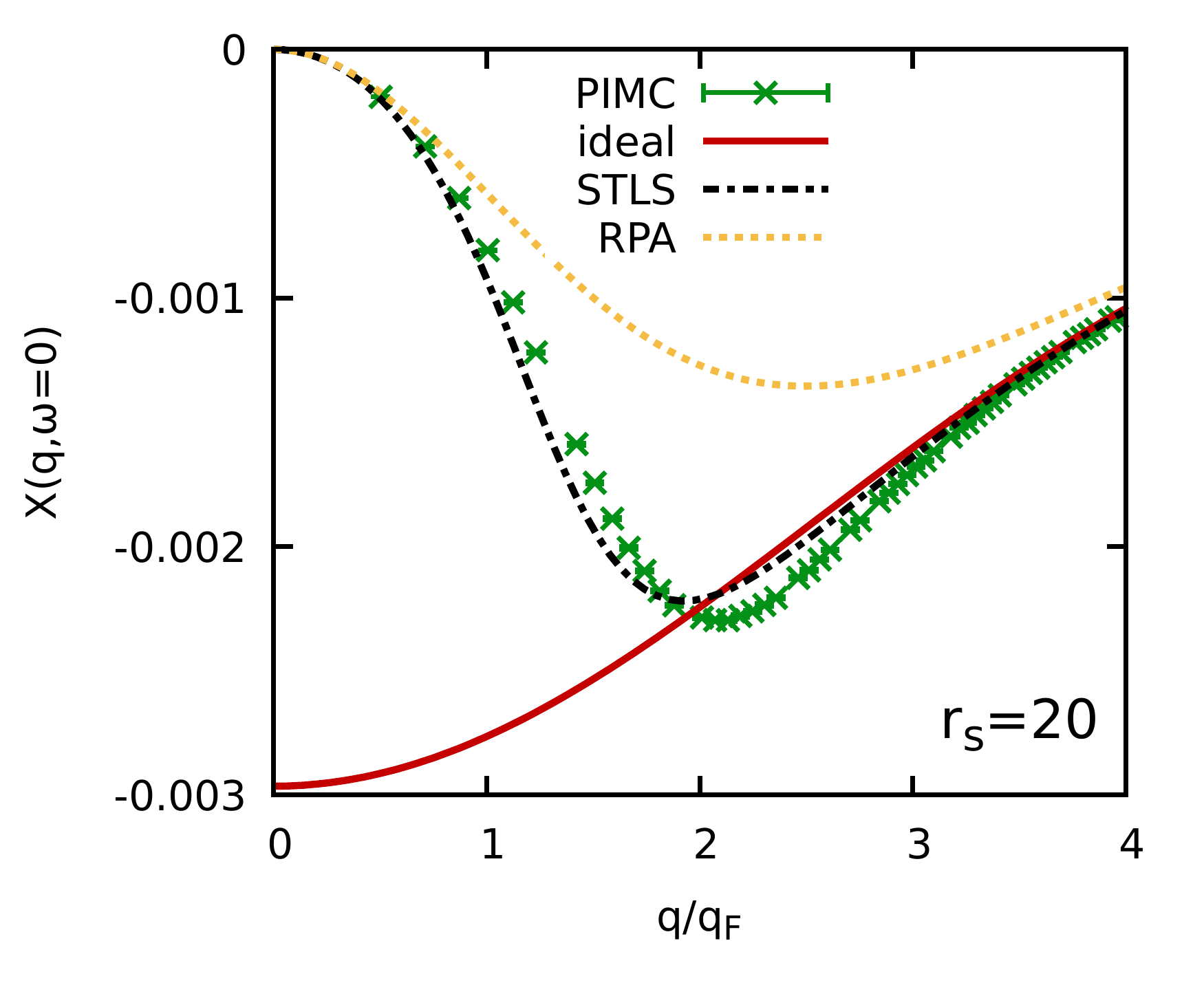} \vspace*{-0.3cm}
\caption{\label{fig:CHI_theta2}
Wave number dependence of the static response function $\chi(q,\omega=0)$ [in Hartree atomic units] at $\theta=2$ for $r_s=4$ (top), $r_s=6$ (center), and $r_s=20$ (bottom). Shown are PIMC data for $N=66$ (green crosses), the ideal function $\chi_0(q,\omega=0)$ (solid red), STLS~\cite{stls,stls2} (dash-dotted black), VS~\cite{stls2} (dashed blue), and RPA (dotted yellow).
}
\end{figure}  
In Fig.~\ref{fig:CHI_theta2}, we show results for the wave number dependence of $\chi(\mathbf{q})$ for $N=66$ unpolarized electrons at $\theta=2$ for three different coupling strengths.
The green crosses depict our PIMC data with the corresponding statistical uncertainty, which is of the order of $\Delta \chi/\chi\sim10^{-3}$ at these parameters. We stress that the evaluation of Eq.~(\ref{eq:static_chi}) allows to evaluate the entire $\mathbf{q}$-dependence of $\chi$ from a single PIMC simulation, which is in stark contrast to the recently proposed simulation of a harmonically perturbed system~\cite{dornheim_pre,groth_jcp}. In the latter case, one has to perform multiple full QMC simulations of an inhomogeneous system governed by Eq.~(\ref{eq:H_perturbed}) for several values of the perturbation amplitude $A$ to obtain $\chi(\mathbf{q})$ for a single wave number. Hence, that strategy is only advisable when $F(\mathbf{q},\tau)$ is not accessible, which is the case for the novel permutation blocking PIMC and configuration PIMC methods that are available over more significant parts of the WDM regime beyond standard PIMC. To summarize, the PIMC evaluation of Eq.~(\ref{eq:static_chi}) constitutes the method of choice wherever possible, and allows to obtain a bulk of accurate data over the entire $q$-range. However, when PIMC breaks down due to the fermion sign problem, the application of PB-PIMC and CPIMC is a valuable alternative of a complementary nature to obtain accurate data for $\chi(\mathbf{q})$ and $G(\mathbf{q})$ in the WDM regime, where quantum degeneracy effects are even more important.

\begin{figure}
\includegraphics[width=0.4147\textwidth]{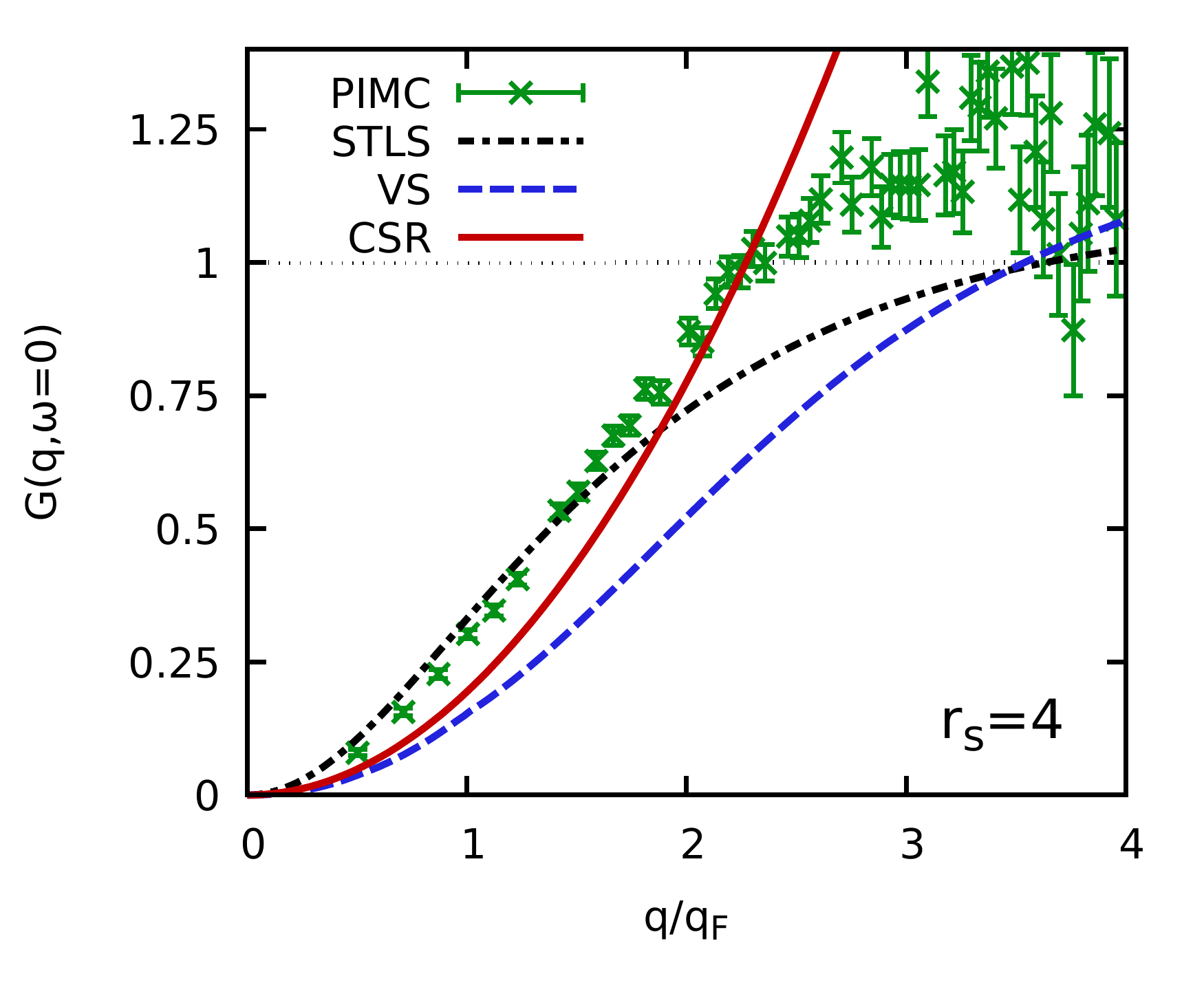}\\ \vspace*{-0.3cm}
\includegraphics[width=0.4147\textwidth]{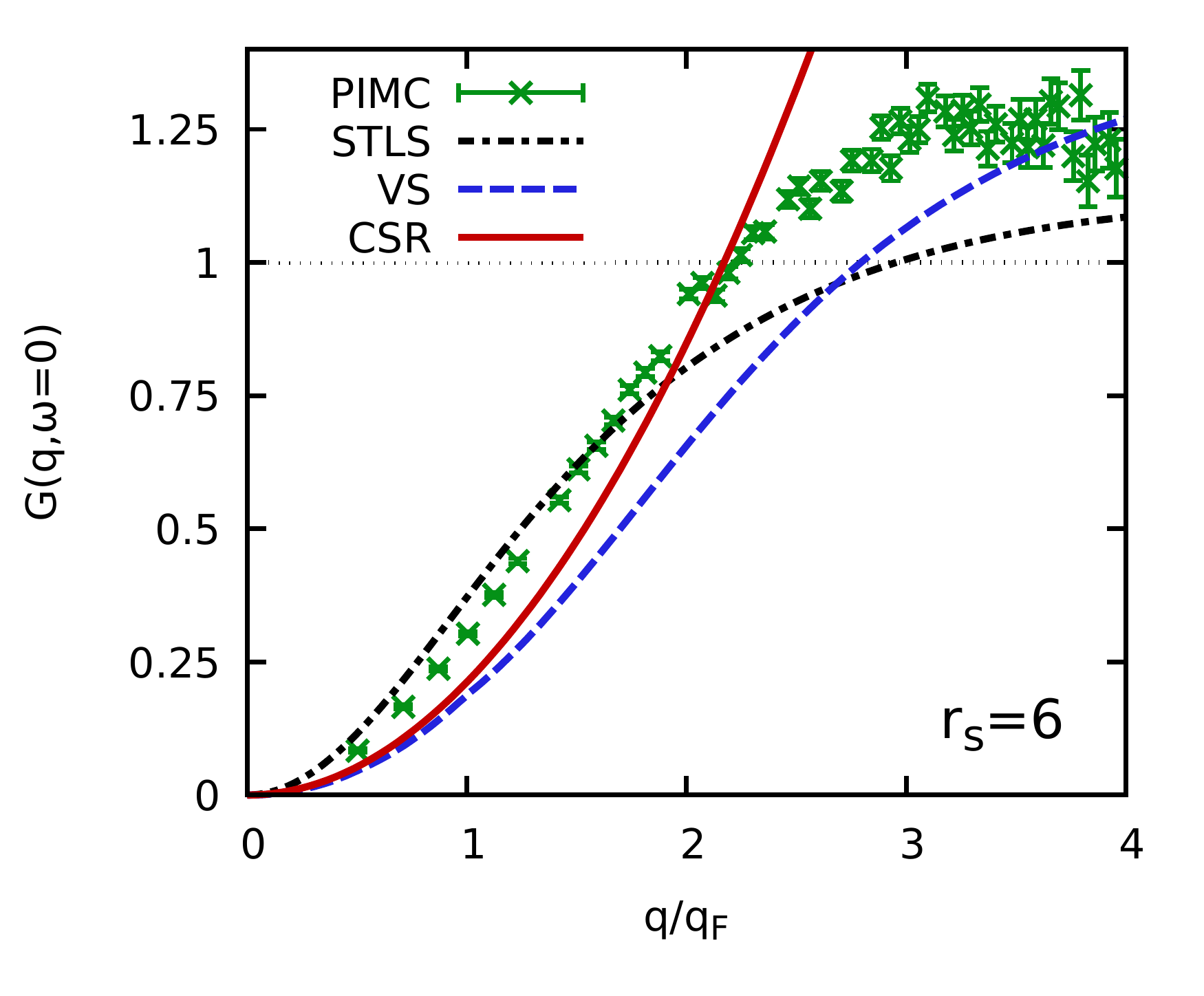}\\ \vspace*{-0.3cm}
\includegraphics[width=0.4147\textwidth]{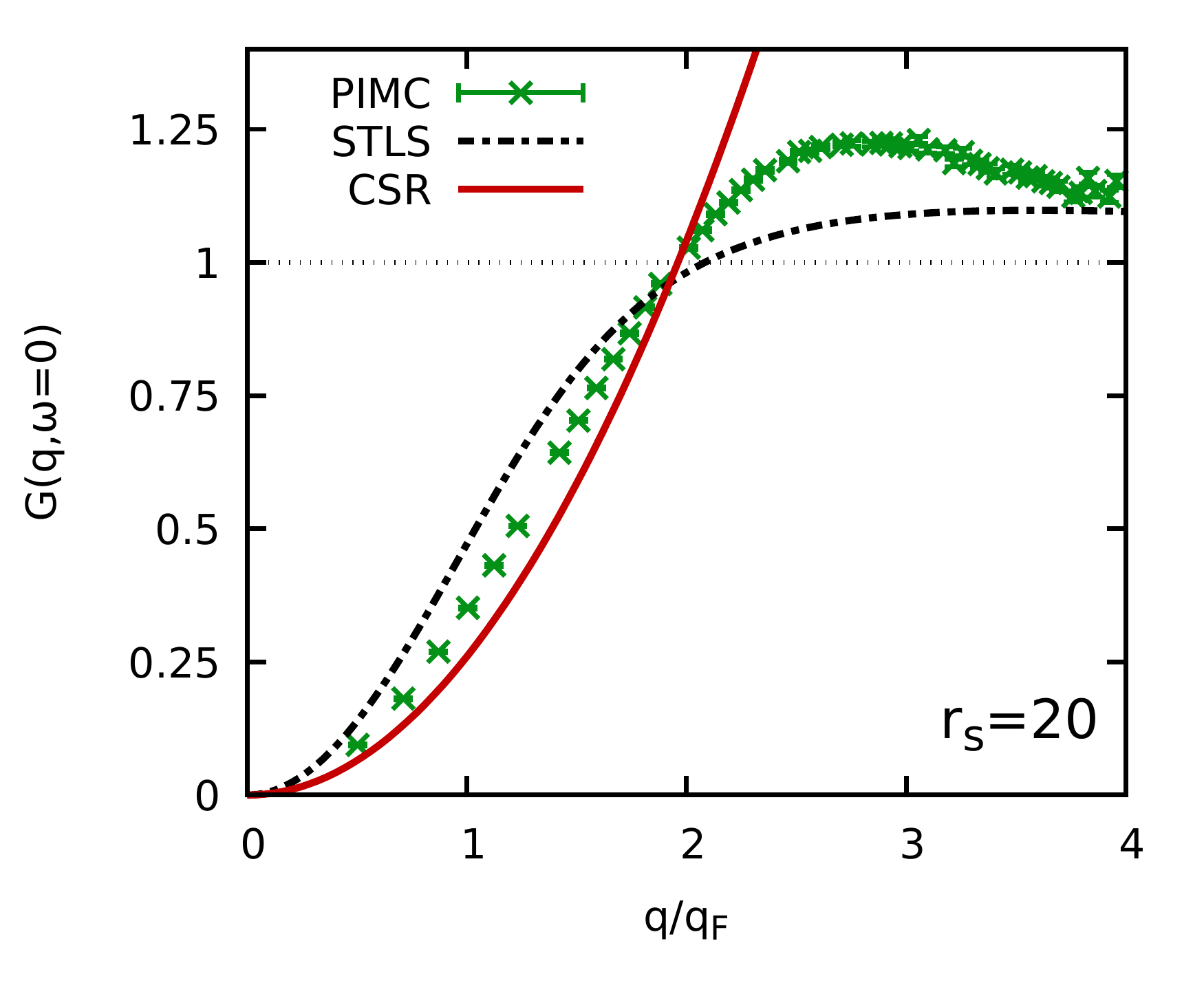} \vspace*{-0.3cm}
\caption{\label{fig:LFC_theta2}
Wave number dependence of the static local field correction $G(q,\omega=0)$ at $\theta=2$ for $r_s=4$ (top), $r_s=6$ (center), and $r_s=20$ (bottom). Shown are PIMC data for $N=66$ and $P=100$ (green crosses),  STLS~\cite{stls,stls2} (dash-dotted black), VS~\cite{stls2} (dashed blue), and the compressibility sum-rule (CSR, solid red), cf.~Eq.~(\ref{eq:csr}).
}
\end{figure}

In the top panel of Fig.~\ref{fig:CHI_theta2}, we compare the PIMC results for $r_s=4$ to the ideal density-response function $\chi_0(\mathbf{q})$ (solid red line) and various other theories, see Ref.~\cite{review} for an extensive recent review article. First and foremost, we note that all data sets (except the ideal curve) exhibit the correct parabolic behavior~\cite{kugler2},
\begin{eqnarray}
\lim_{q\to0}\chi(\mathbf{q}) = - \frac{q^2}{4\pi} \quad ,
\end{eqnarray}
for small wave numbers, which is a direct consequence of perfect screening in the UEG.
The dotted yellow line corresponds to the RPA, which entails a mean-field description of the response to an external perturbation. Thus, the RPA is accurate at high density and temperature, but exhibits only a qualitative agreement with the exact PIMC data even for $r_s=4$, with the systematic deviations being most pronounced for intermediate wave numbers $0.5q_\textnormal{F}\lesssim q \lesssim 3 q_\textnormal{F}$.
The dashed blue curve has been obtained using the approximate static local field correction $G_\textnormal{VS}(\mathbf{q})$ from the Vashishta-Singwi (VS) formalism~\cite{vs_original,stolzmann} taken from Sjostrom and Dufty~\cite{stls2} (see Fig.~\ref{fig:LFC_theta2} for the actual corresponding $G(\mathbf{q})$ data). Evidently, the incorporation of $G_\textnormal{VS}$ leads to a significant improvement over the RPA for all $q$-values, although it remains inaccurate most notably around $q=1.5q_\textnormal{F}$.
Lastly, the dash-dotted black line corresponds to the approximate static local field correction $G_\textnormal{STLS}(\mathbf{q})$ proposed by Singwi, Tosi, Land, and Sj\"olander (STLS)~\cite{stls_original} for the ground-state, which was extended in Refs.~\cite{stls,stls2,tanaka_new} to the WDM regime.
The STLS formalism proves to be significantly more accurate than both RPA and VS, which is in agreement with previous findings for the static structure factor $S(\mathbf{q})$ and the interaction energy $v$ discussed in Refs.~\cite{review,dornheim_cpp,groth_prl,dornheim_pop}.

Upon increasing the coupling strength to $r_s=6$ (center) and $r_s=20$ (bottom), all depicted approximations become noticeably less accurate, although STLS still looks impressive in the former case. However, at strong coupling, even $G_\textnormal{STLS}(\mathbf{q})$ cannot capture the effect of XC contributions and the corresponding $\chi_\textnormal{STLS}(\mathbf{q})$ exhibits both a wrong peak position and magnitude. Nevertheless, the complete negligence of $G$ in the case of RPA leads to systematic deviations of the order of $100\%$.


Let us next discuss the static local field correction $G(\mathbf{q})$, which is shown in Fig.~\ref{fig:LFC_theta2} for the same conditions as $\chi(\mathbf{q})$ in Fig.~\ref{fig:CHI_theta2}. Recall that the RPA corresponds to setting $G_\textnormal{RPA}(\mathbf{q})=0$. In addition, the solid red curve corresponds to the exact limit for small wave numbers that directly follows from the compressibility sum-rule (CSR)~\cite{stls2,review}
\begin{eqnarray}\label{eq:csr}
\lim_{q\to0} G(\mathbf{q}) = - \frac{q^2}{4\pi} \frac{\partial}{\partial n^2} \left(
n f_\textnormal{xc}
\right) \quad ,
\end{eqnarray}
with $n$ being the total electronic density. In practice, Eq.~(\ref{eq:csr}) is evaluated using the accurate parametrization of the XC-free energy from Ref.~\cite{groth_prl}.
Evidently, the CSR-curves do not overlap with our PIMC data (green crosses), as the small-$q$ limit is not accessible due to the finite simulation box, see Refs.~\cite{dornheim_prl,review} for an analog discussion of the static structure factor $S(\mathbf{q})$.
Furthermore, it is well-known that the STLS curve (dash-dotted black) violates the CSR, even if we replace the exact $f_\textnormal{xc}$ in Eq.~(\ref{eq:csr}) by the corresponding STLS expression $f^\textnormal{STLS}_\textnormal{xc}$, see, e.g., Ref.~\cite{stls2}.
In contrast, the VS-curve is in good agreement with the CSR, which is an understandable, yet nontrivial finding: while the expression for the static local field correction $G_\textnormal{VS}(\mathbf{q})$ is derived by imposing the CSR, this is done by assuming the approximate expression for $f_\text{xc}$ within the VS formalism, $f_\textnormal{xc}^\textnormal{VS}$, which is known to be even less accurate than, for example, $f_\textnormal{xc}^\textnormal{STLS}$. Still, this condition leads to an accurate static local field correction for small $q$.

For larger wave numbers, we observe an increased statistical uncertainty, which is a direct consequence of the evaluation of Eq.~(\ref{eq:static_chi}): within a PIMC simulation, we obtain data for $F(\mathbf{q},\tau)$, which, in turn, is used to compute the static density-response function $\chi(\mathbf{q})$ via Eq.~(\ref{eq:chi_static}). The static local field correction is then obtained by measuring the deviation of $\chi$ to the noninteracting case, cf.~Eq.~(\ref{eq:G_static}). For large wave numbers $q$, the effect of $G(\mathbf{q})$ on the total response function $\chi$ is suppressed by the $4\pi/q^2$ pre-factor in Eq.~(\ref{eq:define_LFC}) and $\chi$ eventually converges towards the ideal function $\chi_0(\mathbf{q})$, see Fig.~\ref{fig:CHI_theta2}. Therefore, $G$ becomes the relatively small difference between two large numbers in this regime, and the relative uncertainty in this quantity increases. Still, we stress that the quality of the present QMC data for $G(\mathbf{q})$ is significantly better than analog results from simulations of the harmonically perturbed system both at finite temperature~\cite{dornheim_pre,groth_jcp} and in the ground state~\cite{moroni2,bowen2}.

Despite the relatively high accuracy of the VS and, in particular, the STLS scheme in the description of $\chi(\mathbf{q})$ at $r_s=4$ and $r_s=6$, see Fig.~\ref{fig:CHI_theta2}, these approximations perform significantly worse for $G(\mathbf{q})$ itself. This is not surprising, as the static local field correction provides a wave-number resolved description of exchange-correlation effects and, therefore, is even more sensitive to an approximate treatment of the Coulomb repulsion than $f_\textnormal{xc}$. Furthermore, both VS and STLS are static theories, which means that they completely neglect the frequency-dependence of $G(\mathbf{q},\omega)$. Naturally, this shortcoming affects the static limit of this quantity, which, in turn means that such theories cannot be systematically improved to provide exact results for $G(\mathbf{q})$. In contrast, while our PIMC simulations, too, only yield results for the static limit of $G$, the frequency-dependence of XC-effects is fully included in the imaginary-time formulation of the approach and the corresponding PIMC results are exact.

Finally, in the bottom panel of Fig.~\ref{fig:LFC_theta2} we show results for $G$ for strong coupling, $r_s=20$. In this case, the STLS scheme can only provide a qualitative description and the maximum in $G$ around $q=2.5q_\textnormal{F}$ is not captured at all.

\begin{figure}
\includegraphics[width=0.4147\textwidth]{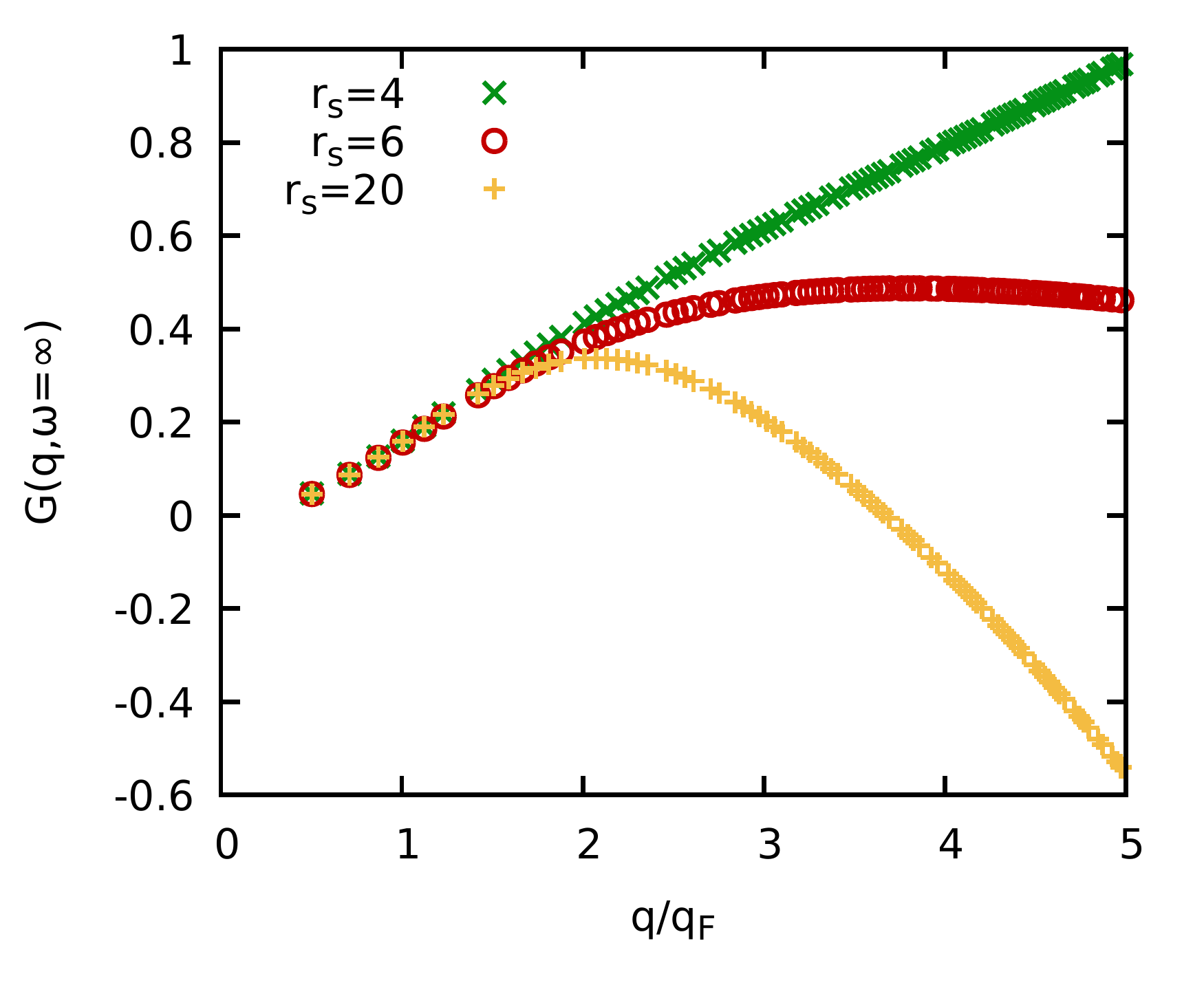}\\
\includegraphics[width=0.4147\textwidth]{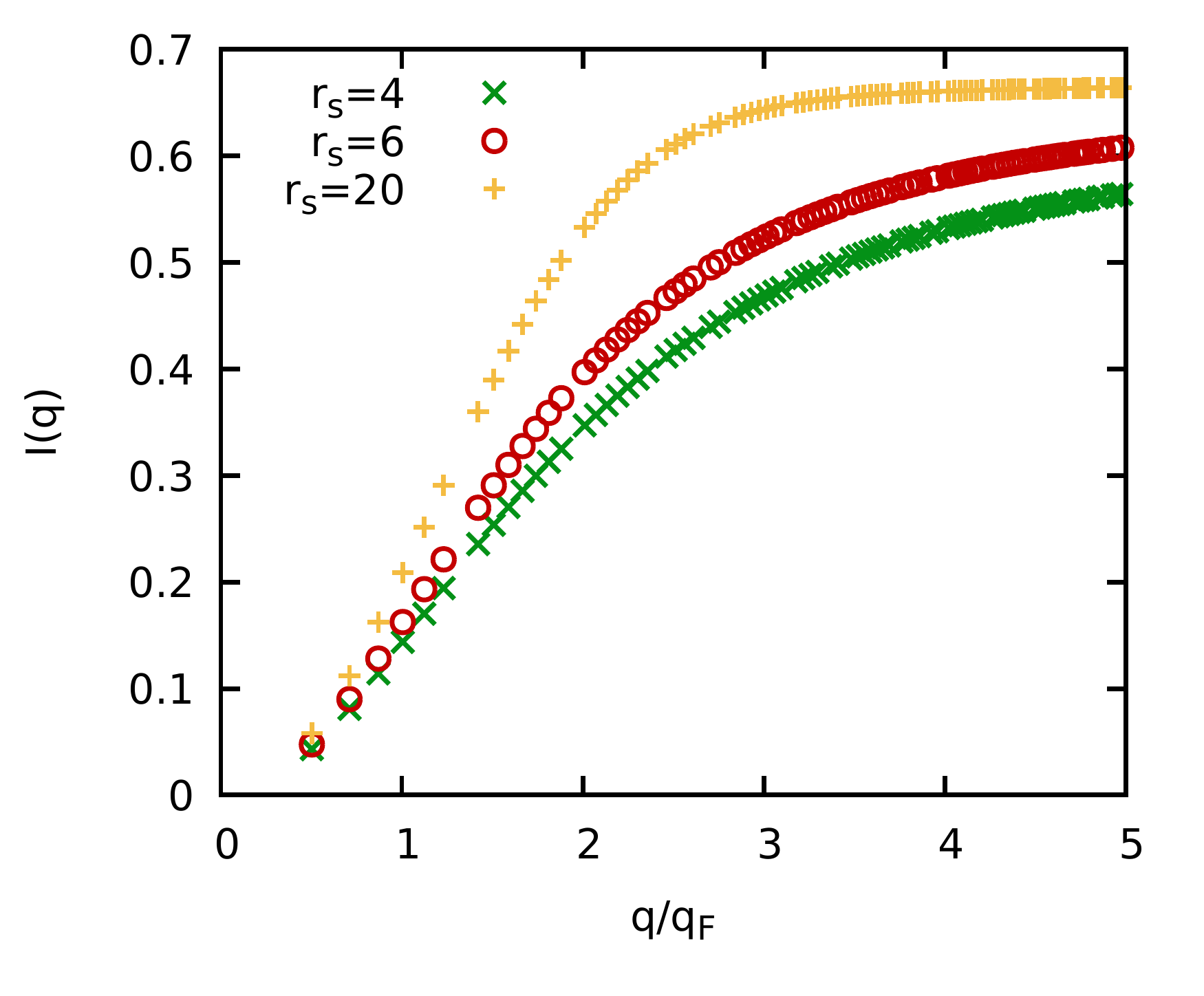}
\caption{\label{fig:G_infinity_theta2_rs}
Wave number dependence of the high-frequency limit of the dynamic local field correction $G(q,\omega=\infty)$ (top) and interaction integral $I(q)$ (bottom, Hartree atomic units) at $\theta=2$ and $N=66$ for $r_s=4$ (green crosses), $r_s=6$ (red circles), and $r_s=20$ (yellow pluses). 
}
\end{figure}

Lastly, let us consider the high-frequency limit of the local field correction, which is obtained from our PIMC data for $S(\mathbf{q})$ and the kinetic energy $K$ via Eq.~(\ref{eq:G_infty}).
\begin{figure}
\includegraphics[width=0.4147\textwidth]{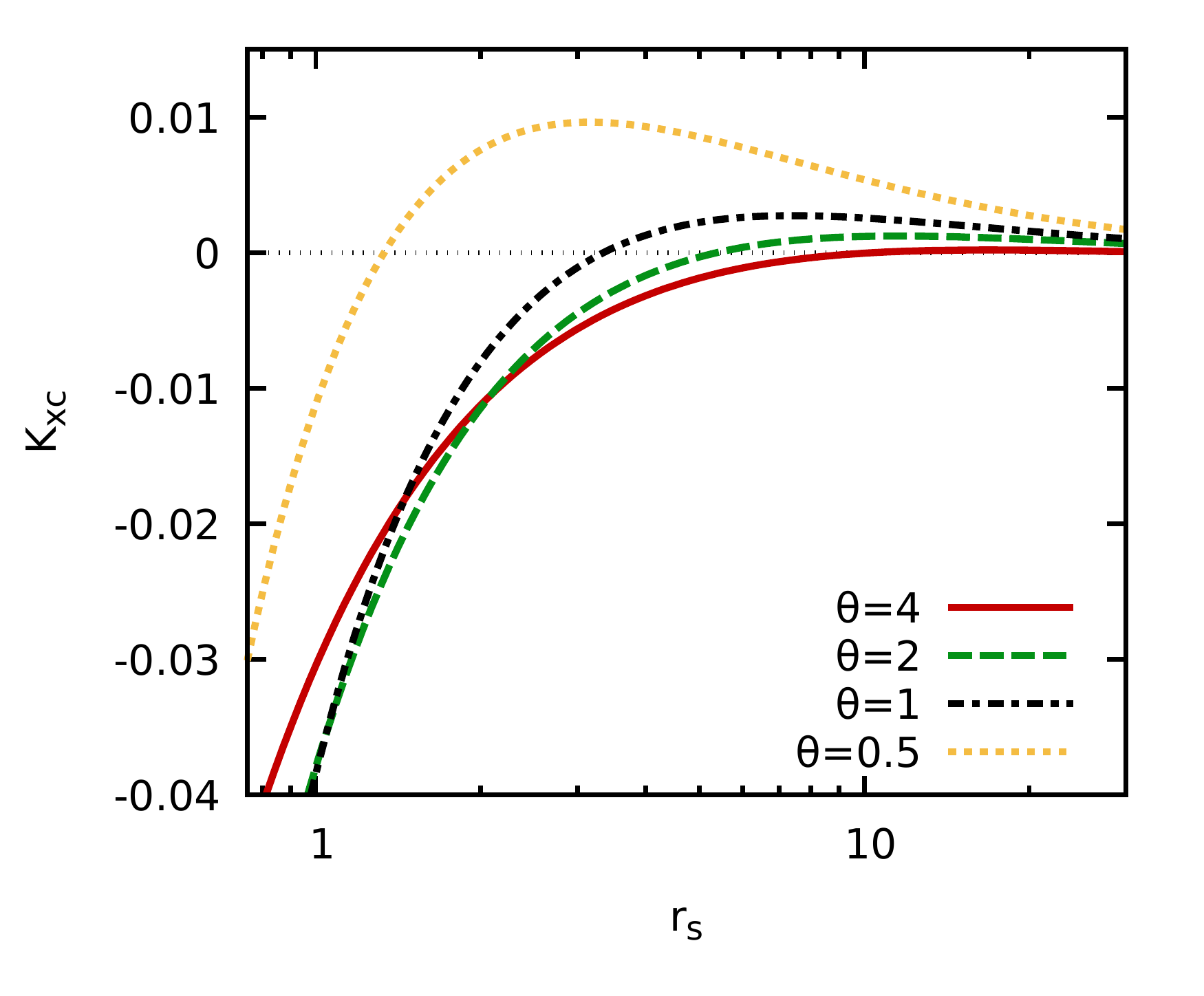}
\caption{\label{fig:Kxc}
Density dependence of the exchange--correlation contribution to the kinetic energy $K_\textnormal{xc}$ [in Hartree atomic units] for $\theta=4$ (solid red), $\theta=2$ (dashed green), $\theta=1$ (dash-dotted black), and $\theta=0.5$ (dotted yellow). Computed from the accurate parametrization of $f_\textnormal{xc}$ from Ref.~\cite{groth_prl}.
}
\end{figure}  
In the top panel of Fig.~\ref{fig:G_infinity_theta2_rs}, we show results for the wave number dependence of $G(\mathbf{q},\omega=\infty)$ for the same conditions as in Fig.~\ref{fig:LFC_theta2}.
Evidently, the three data sets for $r_s=4$ (green crosses), $r_s=6$ (red circles), and $r_s=20$ (yellow pluses) exhibit a quite distinct behavior: while all curves are in good agreement for small $q$, the $r_s=4$ results for $G(\mathbf{q},\omega=\infty)$ continue to increase for large $q$, whereas both the $r_s=6$ curve and, to an even more significant degree, the $r_s=20$ curve decrease in this regime. In fact, the yellow curve attains negative values for $q\gtrsim 3q_\textnormal{F}$. This finding can be understood by considering the two contributions to $G(\mathbf{q},\omega=\infty)$, i.e., the interaction integral $I(q)$ [see Eq.~(\ref{eq:I})] and a parabola with a pre-factor proportional to the negative XC-part to the kinetic energy $-K_\textnormal{xc}$ [see Eq.~(\ref{eq:Kxc})]. For small wave numbers, $I(q)$ dominates as can be seen in the bottom panel of Fig.~\ref{fig:G_infinity_theta2_rs} and is monotonically increasing with $q$. Eventually, however, the parabolic term takes over, and the particular behavior of $G(\mathbf{q},\omega=\infty)$ is determined by $K_\textnormal{xc}$, which is plotted in Fig.~\ref{fig:Kxc}.
\begin{figure}
\includegraphics[width=0.4147\textwidth]{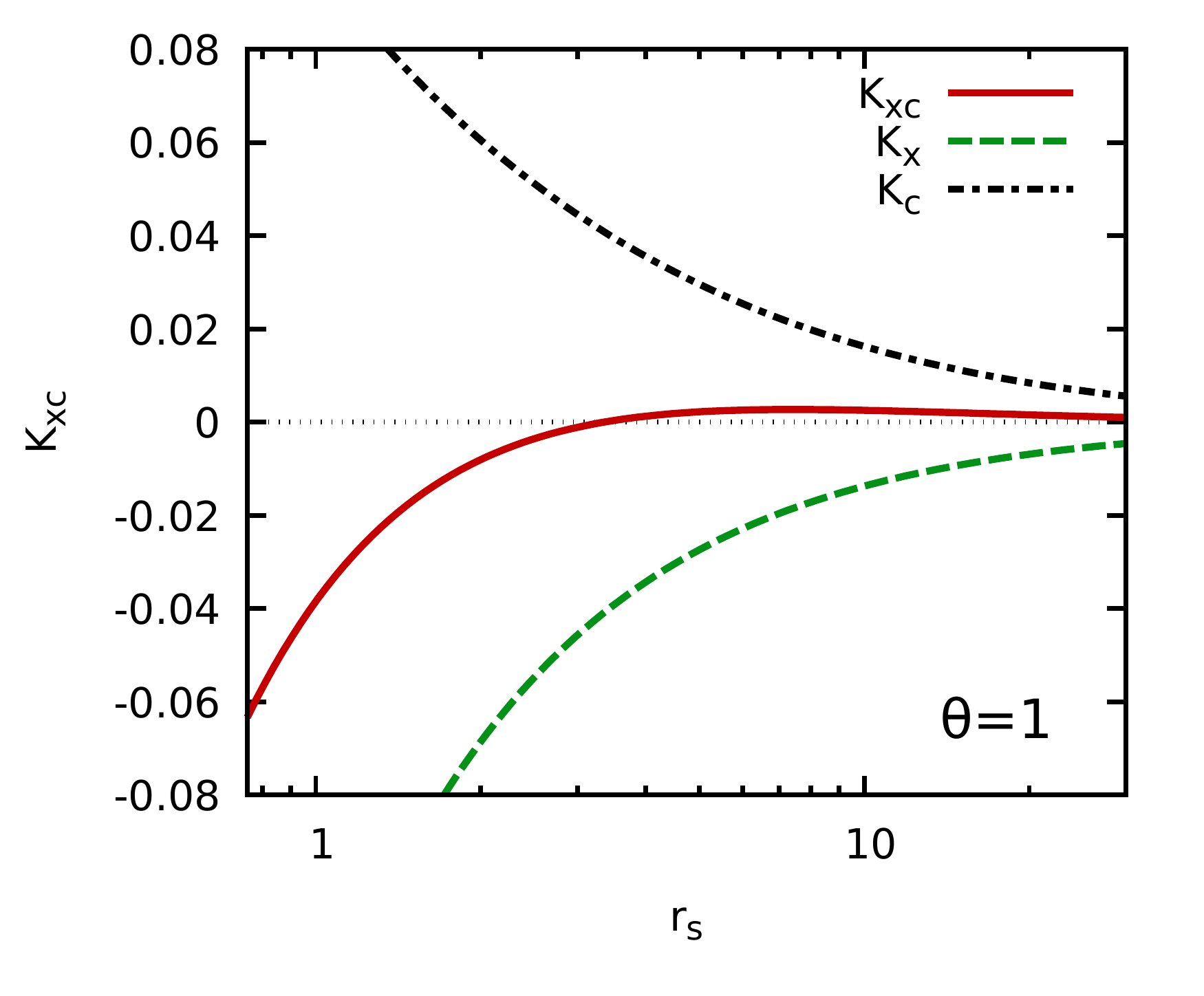}\\
\includegraphics[width=0.4147\textwidth]{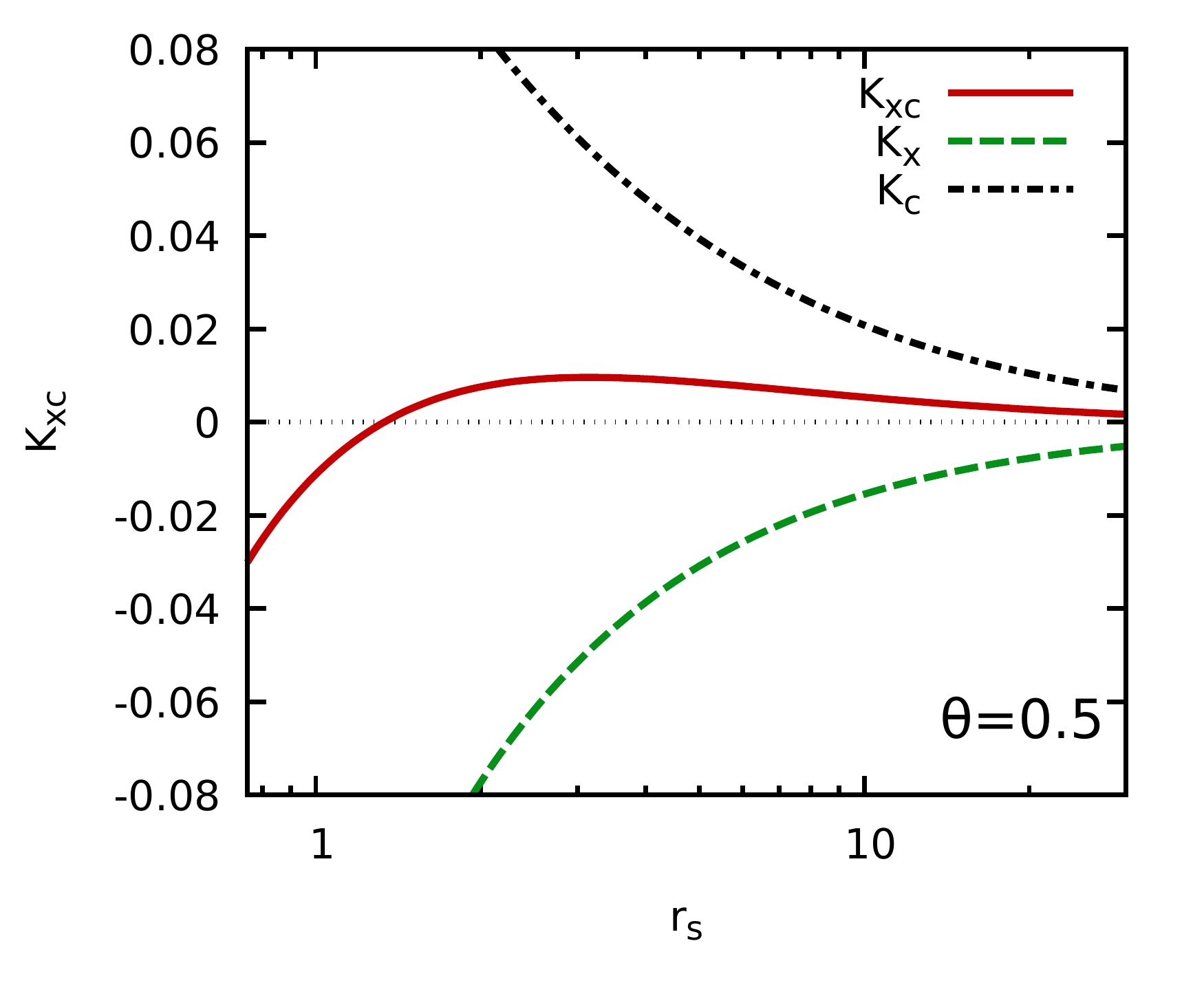}
\caption{\label{fig:Kxc_contributions}
Shown is the density dependence of the exchange--correlation kinetic energy $K_\textnormal{xc}$ [in Hartree atomic units], and the separate contributions $K_\textnormal{x}$ and $K_\textnormal{c}$, for $\theta=1$ (top) and $\theta=0.5$ (bottom). Computed from the accurate parametrization of $f_\textnormal{xc}$ from Ref.~\cite{groth_prl} and of $f_\textnormal{x}$ from Ref.~\cite{F}.
}
\end{figure}  
For the highest depicted reduced temperature, $\theta=4$ (solid red), $K_\textnormal{xc}$ is negative for $r_s\lesssim10$. In contrast, for $\theta=0.5$ (dotted yellow) this is the case for $r_s\lesssim1$. With these findings, the observed large-$q$ behavior in the top panel of Fig.~\ref{fig:G_infinity_theta2_rs} becomes obvious: negative values of $K_\text{xc}$, as in the case of $\theta=2$ and $r_s=4$, lead to positive pre-factors of the parabola and $G(\mathbf{q},\omega=\infty)$ remains strictly increasing; positive values of $K_\textnormal{xc}$, on the other hand, result in a negative parabola as observed for $r_s=6$ and $r_s=20$. For completeness, we note that similar findings have been observed for the ground state in Ref.~\cite{iwamoto}.

Let us conclude this section by investigating the origin of the nontrivial progression of $K_\textnormal{xc}$.
To this end, we plot the individual exchange and correlation parts, $K_\textnormal{x}$ (dashed green) and $K_\textnormal{c}$ (dash-dotted black), to $K_\textnormal{xc}$ (solid red) in Fig.~\ref{fig:Kxc_contributions}. Note that $K_\textnormal{x}$ is obtained by replacing $f_\textnormal{xc}$ in Eq.~(\ref{eq:Kxc}) with the noninteracing free energy $f_\textnormal{x}$, which was parametrized by Perrot and Dharma-wardana~\cite{F}. Interestingly, $K_\textnormal{xc}$ is given by the relatively small difference of the positive exchange and the negative correlation parts. More specifically, the correlation contribution dominates for low temperatures and large values of the coupling parameter $r_s$, and $K_\textnormal{xc}$ attains positive values. Evidently, this happens for even smaller $r_s$-values at the lower temperature, $\theta=0.5$ (bottom panel).

\subsection{Stochastic sampling method\label{sec:stochastic_sampling_results}}

\begin{figure}
\includegraphics[width=0.4647\textwidth]{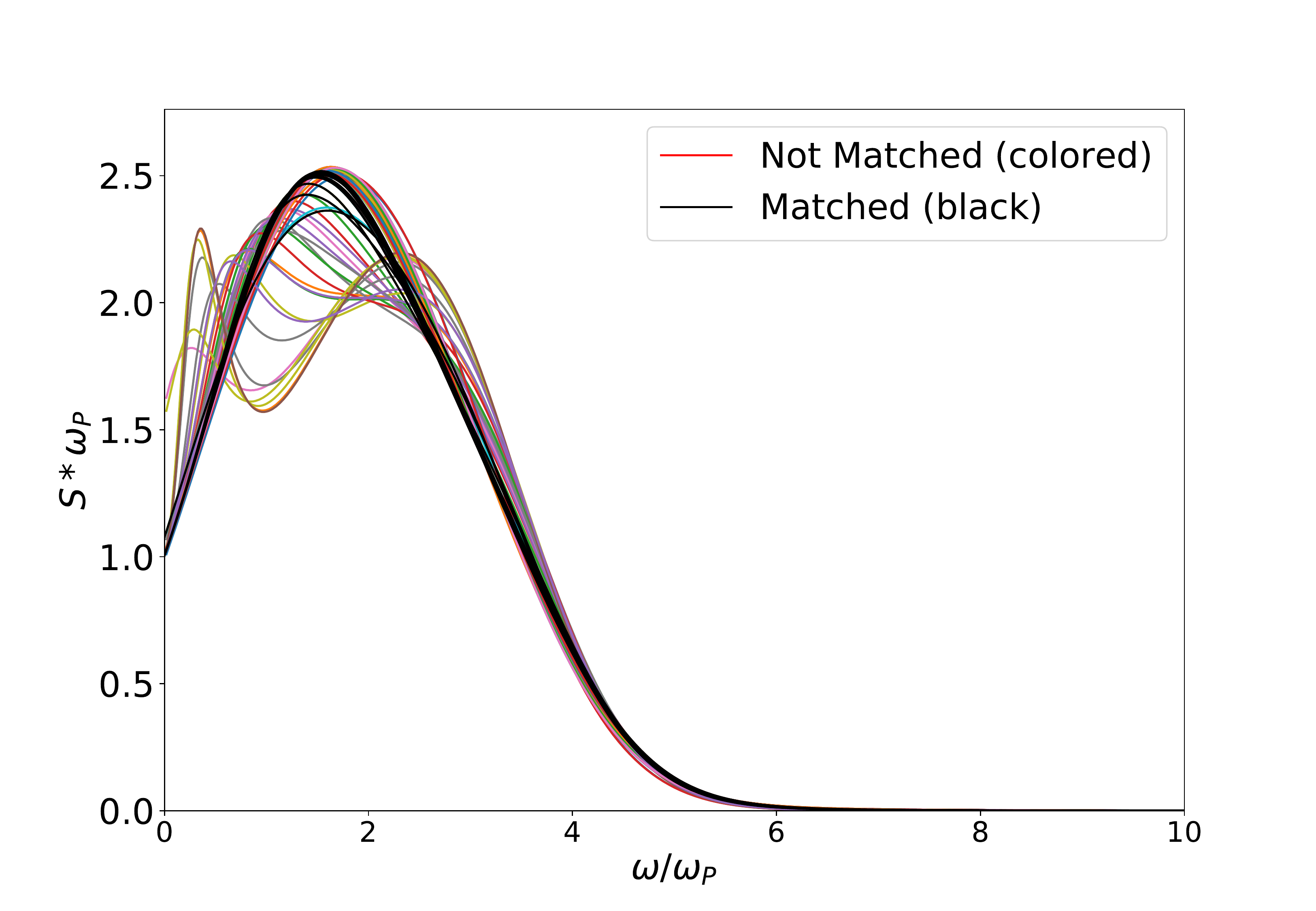}\\ \vspace*{-0.45cm}
\includegraphics[width=0.4647\textwidth]{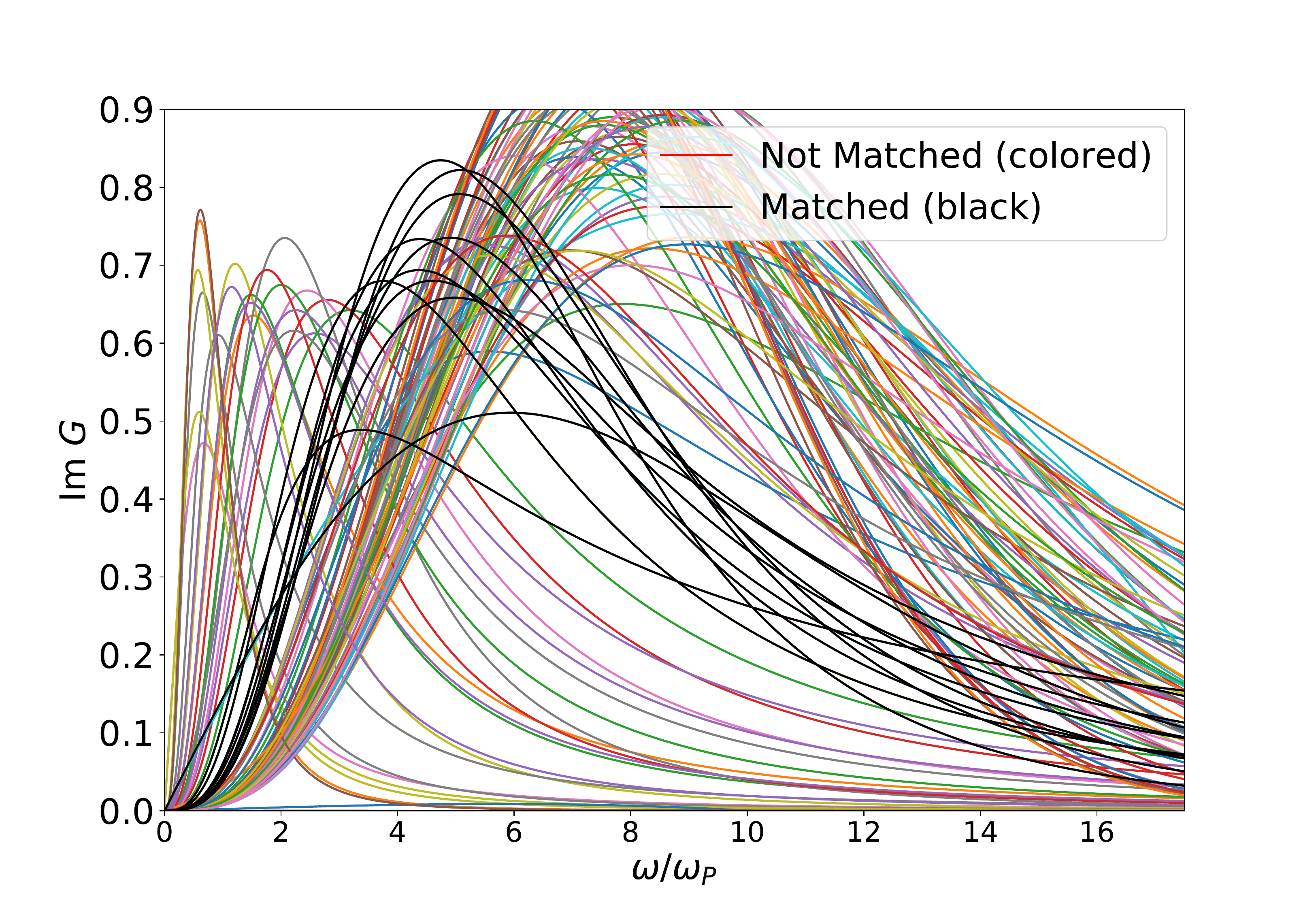}\\\vspace*{-0.45cm}
\includegraphics[width=0.4647\textwidth]{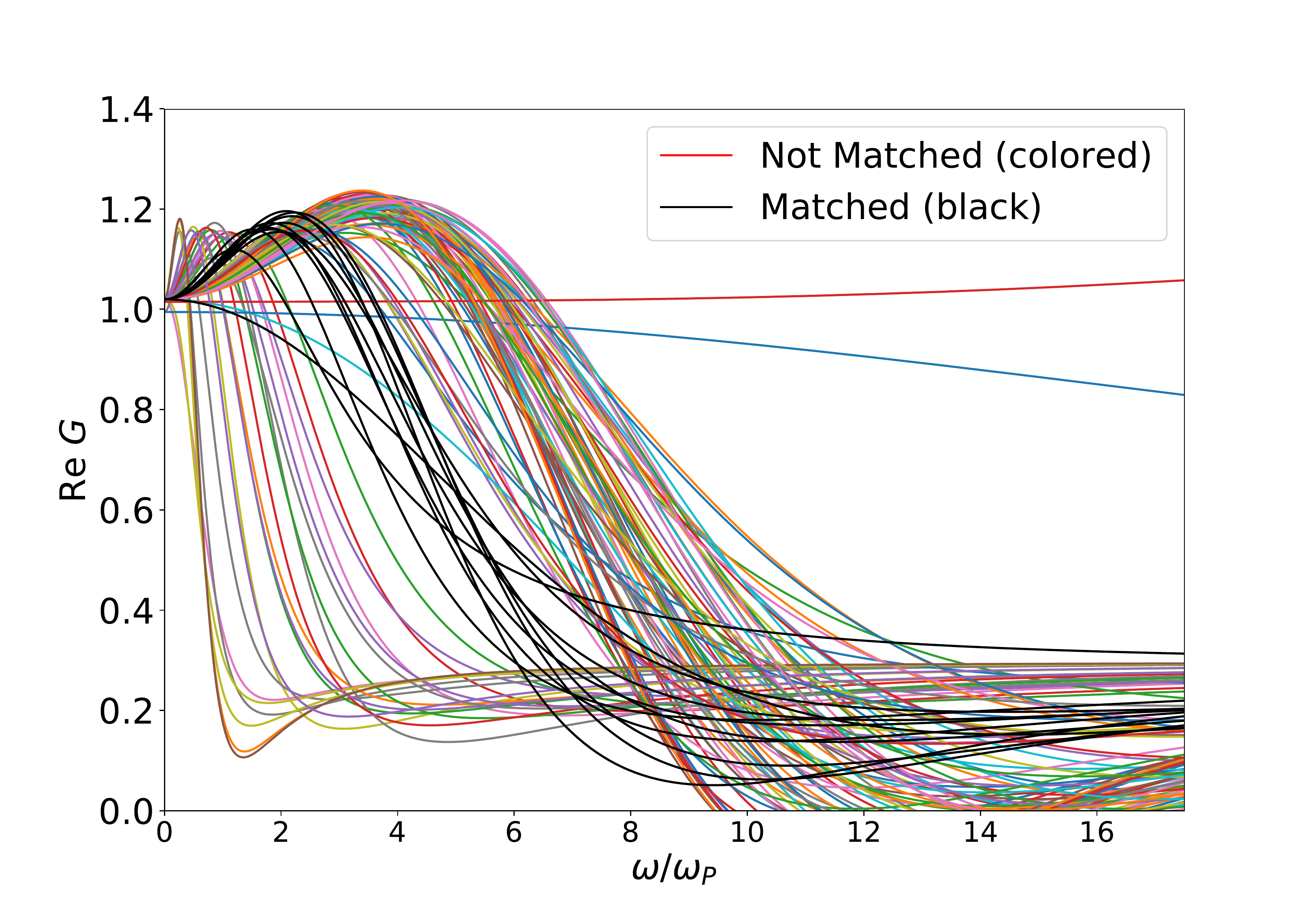}
\caption{\label{fig:example_theta1}
Reconstruction of the dynamic structure factor $S(|\mathbf{q}|\approx1.98q_\textnormal{F},\omega)$ for $N=34$ unpolarized electrons at $r_s=6$ and $\theta=1$. The top panel shows a set of $M=115$ trial solution $S_{\textnormal{trial},i}$, with the black (coloured) curves having (not) been included into the final average, Eq.~(\ref{eq:final_average}). The central and bottom panels depict the corresponding trial functions for the imaginary and real parts of the dynamic local field correction, see Sec.~\ref{sec:stochastic_sampling} for an extensive theoretical introduction.
}
\end{figure}

In the following section, we give a hands-on discussion of two examples of our stochastic sampling scheme (see Sec.~\ref{sec:stochastic_sampling}) to compute the dynamic structure factor $S(\mathbf{q},\omega)$ from our PIMC data. See Sec.~\ref{sec:dynamic_results}, for a discussion of our findings for this quantity for different densities and temperatures.

\begin{figure}
\includegraphics[width=0.4647\textwidth]{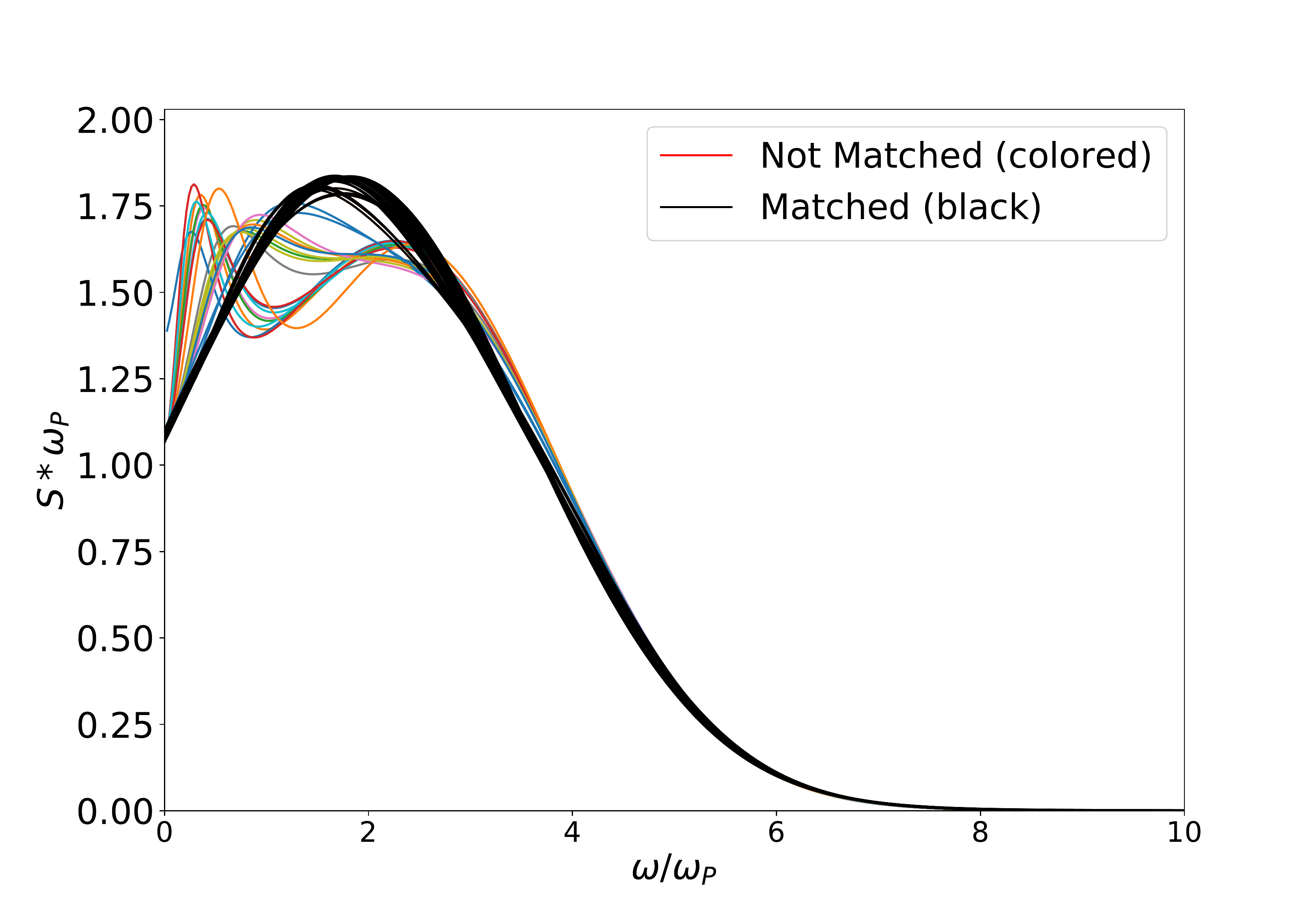}\\ \vspace*{-0.45cm}
\includegraphics[width=0.4647\textwidth]{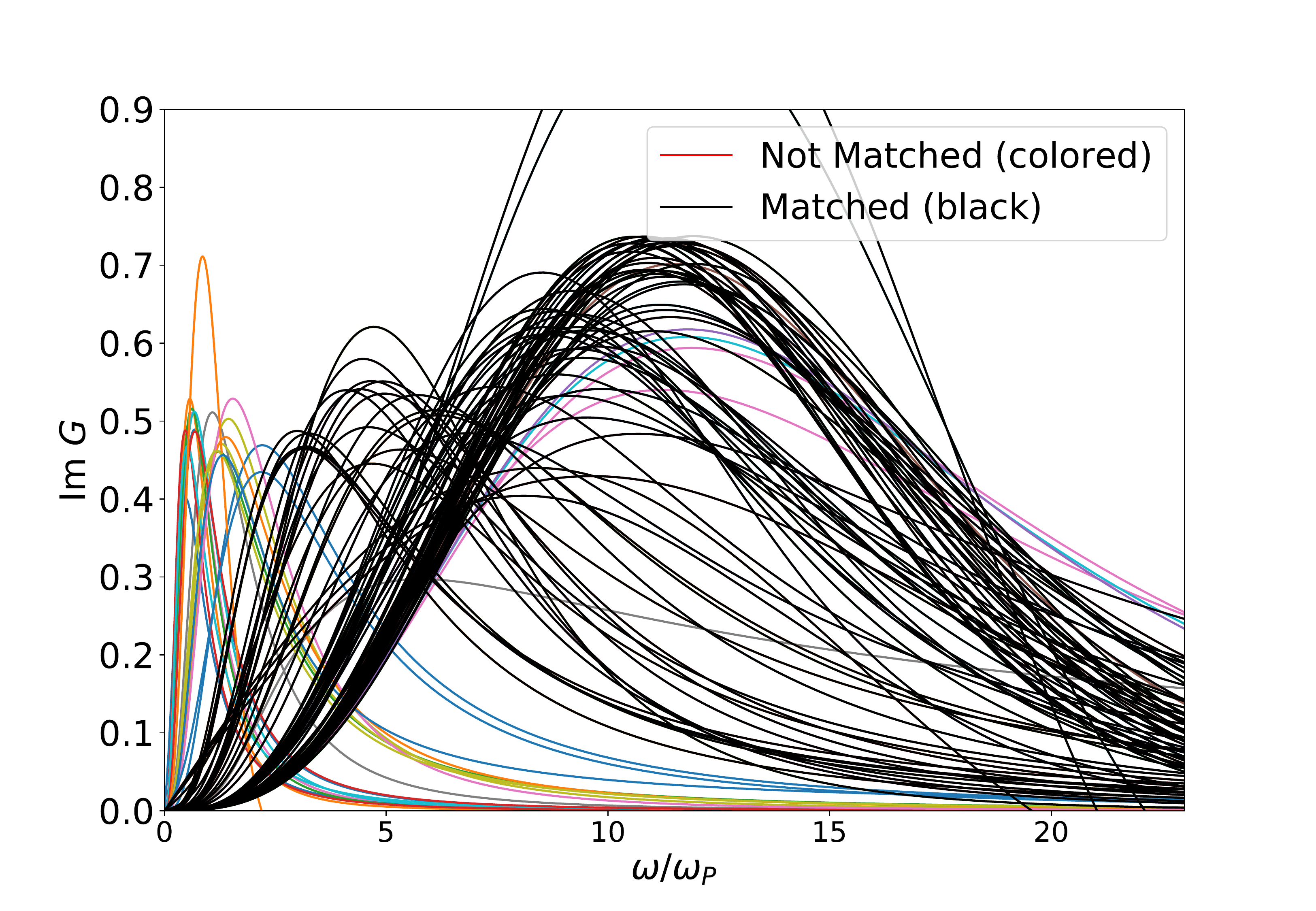}\\\vspace*{-0.45cm}
\includegraphics[width=0.4647\textwidth]{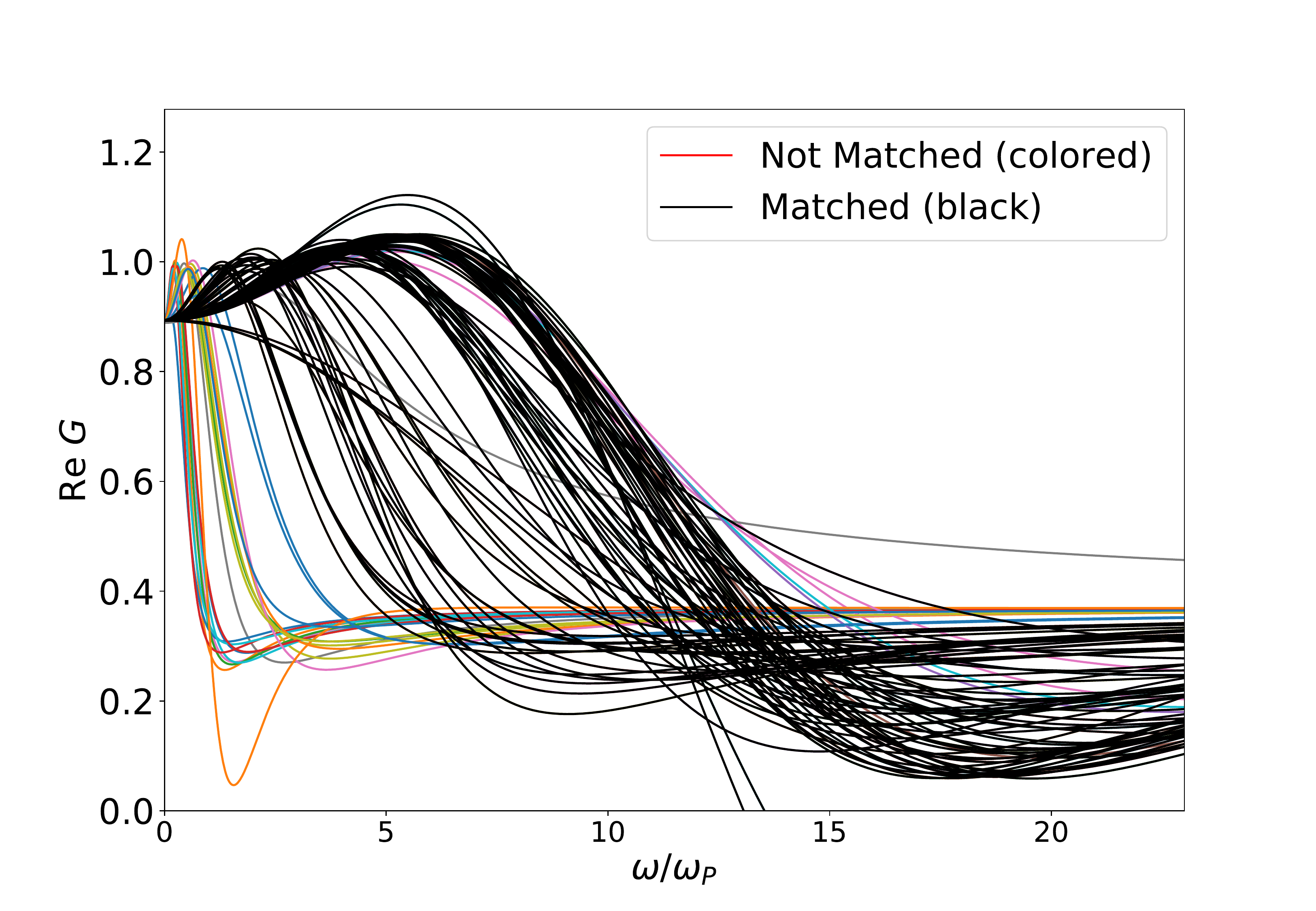}
\caption{\label{fig:example_theta2}
Reconstruction of the dynamic structure factor $S(|\mathbf{q}|\approx1.98q_\textnormal{F},\omega)$ for $N=34$ unpolarized electrons at $r_s=6$ and $\theta=2$. The top panel shows a set of $M=99$ trial solution $S_{\textnormal{trial},i}$, with the black (coloured) curves having (not) been included into the final average, Eq.~(\ref{eq:final_average}). The central and bottom panels depict the corresponding trial functions for the imaginary and real parts of the dynamic local field correction, see Sec.~\ref{sec:stochastic_sampling} for an extensive theoretical introduction.
}
\end{figure}

In Fig.~\ref{fig:example_theta1}, we show trial solutions for $S(\mathbf{q},\omega)$ (top), $\textnormal{Im} G(\mathbf{q},\omega)$ (center), and $\textnormal{Re} G(\mathbf{q},\omega)$ (bottom) that have been generated by following the procedure from Sec.~\ref{sec:stochastic_sampling} for $N=34$ electrons at $r_s=6$ and $\theta=1$ around twice the Fermi wave vector. In particular, the black curves are \emph{valid} solutions that are included in the final average $S_\textnormal{final}(\mathbf{q},\omega)$ [Eq.~(\ref{eq:final_average})], whereas the coloured curves violate the corresponding imaginary-time correlation--function $F(\mathbf{q},\tau)$ and the frequency moments $\braket{\omega^k}$.
Upon examining the dynamic structure factor, we find that the stochastic sampling of $G(\mathbf{q},\omega)$ leads to (at least) two distinct classes of trial solution for $S$. More specifically, there appear $S_\textnormal{trial}(\mathbf{q},\omega)$ with a pronounced double-peak structure (which do not fulfill the PIMC boundary conditions), with the first one sometimes being interpreted as a \emph{diffusive peak}~\cite{jan1}. In addition, there are solutions with a single broad peak around $\omega=2\omega_\textnormal{p}$, which provide those $S$ that fulfill all known requirements from our PIMC simulations. Interestingly, though, slight variations in the form and/or position of the peak lead to a significant disagreement regarding $F(\mathbf{q},\tau)$ and $\braket{\omega^k}$, and the corresponding $S_\textnormal{trial}$ are, consequently, discarded. This is in stark contrast to the $\theta=2$ case, which is show in Fig.~\ref{fig:example_theta2} and discussed below.

The central panel of Fig.~\ref{fig:example_theta1} shows the corresponding stochastically sampled imaginary parts of the dynamic local field correction, cf.~Eq.~(\ref{eq:parametrization}). All depicted curves exhibit a fairly similar progression with the exact conditions $\textnormal{Im}G(\mathbf{q},0) = \textnormal{Im}G(\mathbf{q},\infty)=0$ and a single peak in between. However, the peak position, width, and high-frequency tail are significantly different from each other. Moreover, the bottom panel shows the real part of $G(\mathbf{q},\omega)$, which has been obtained from $\textnormal{Im} G(\mathbf{q},\omega)$ via the Kramers-Kronig relation Eq.~(\ref{eq:Kramers_Kronig_real}) by numerical integration. Interestingly, this procedure leads to a more complicated functional form: starting from the exact static limit for $\omega\to 0$, $\textnormal{Re}G (\mathbf{q},\omega)$ always exhibits a monotonic increase for small $\omega$, followed by a maximum, which varies drastically in its position depending on the chosen trial parameters $a_i$, $b_i$, and $c$. Moreover, there appears a minimum for intermediate frequencies, until the exact high-frequency limit is reached from below. Again, the particular form of this tail can be very different.

While the mapping from $G_\textnormal{trial}(\mathbf{q},\omega)$ onto $S_\textnormal{trial}(\mathbf{q},\omega)$ is far from obvious, we find that those $\textnormal{Im}G (\mathbf{q},\omega)$ with a maximum at low frequency result in a double-peaked dynamic structure factor with the aforementioned diffusive feature. In addition, we note that fairly different trial solutions both in the imaginary and real parts of the dynamic local field correction lead to very similar results in $S_\textnormal{trial}(\mathbf{q},\omega)$ itself. Since our PIMC data for $F(\mathbf{q},\tau)$ and $\braket{\omega^k}$ only allow us to decide between differences in the dynamic structure factor itself, it follows that $G(\mathbf{q},\omega)$ is much less restrained by our reconstruction procedure than $S(\mathbf{q},\omega)$.

As a side remark, we mention that the parametrization of $\textnormal{Im}G (\mathbf{q},\omega)$ from Eq.~(\ref{eq:parametrization}) has proven sufficiently flexible even at strong coupling ($r_s=10,20$), when a nontrivial incipient excitonic feature emerges. Still, it can potentially be systematically improved by including higher powers in $\omega$, which will likely be necessary for even stronger coupling, which, however, is beyond the scope of the present work. 
Further, we mention that only $7$ of the $M=115$ trial solution shown in Fig.~\ref{fig:example_theta1} are actually included into the computation of the final result, and the distribution of the $\textnormal{Re}G$ and $\textnormal{Im}G$ curves is far from uniform. This leaves open the possibility to further improve our sampling procedure in future work. Moreover, the parametrization of $\textnormal{Im}G$ and subsequent calculation of $\textnormal{Re}G$ (cf.~Sec.~\ref{sec:stochastic_sampling}) that is employed in this work is not the only choice, and other interpolations between the known limits of $G(\mathbf{q},\omega)$ have been reported elsewhere~\cite{jan1,hong1,hong2}.

Let us conclude this discussion of the stochastic sampling reconstruction procedure by increasing the temperature to $\theta=2$, which is shown in Fig.~\ref{fig:example_theta2}. Remarkably, in this case we find a significantly increased ratio of included (black) to discarded (coloured) solutions. More specifically, the appearing double-peak structure factors are still not in agreement with our PIMC data for $F(\mathbf{q},\tau)$ and $\braket{\omega^k}$, whereas nearly all solutions from the class with only a single broad peak are included into the construction of the final average $S_\textnormal{final}(\mathbf{q},\omega)$.
Upon examining the corresponding dynamic local field corrections, we find that almost only those $\textnormal{Im}G (\mathbf{q},\omega)$ are discarded that have a maximum at very low-frequency. Consequently, $G(\mathbf{q},\omega)$ is even less determined by our PIMC data for the higher temperature than for $\theta=1$, as the impact on observable quantities is smaller. Still, the dynamic structure factor itself remains of high quality.

\subsection{Dynamic structure factors\label{sec:dynamic_results}}

Let us conclude this paper with a presentation of new results for the dynamic structure factor $S(\mathbf{q},\omega)$ for different densities and temperatures.

\begin{figure}
\includegraphics[width=0.4647\textwidth]{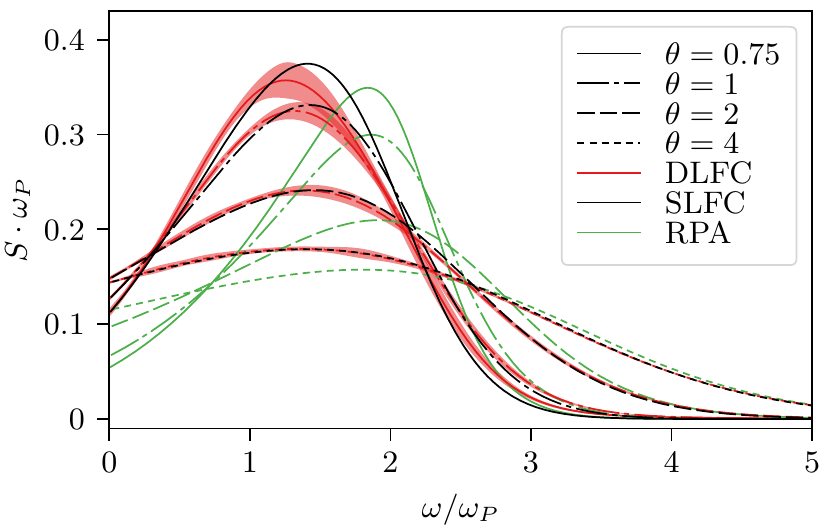}
\caption{\label{fig:TTT}
Dynamic structure factor $S(q\approx1.4q_\textnormal{F},\omega)$ of the warm dense electron gas with $N=34$ and $r_s=6$ for different temperatures. The solid red, dashed black, and solid green lines depict data from our stochastically sampled dynamic local field correction, the static approximation $G_\textnormal{static}(\mathbf{q},\omega)=\textnormal{Re}G(\mathbf{q},0)$, and the random phase approximation (RPA), respectively.
}
\end{figure}

In Fig.~\ref{fig:TTT}, we show the dynamic structure factor $S(\mathbf{q},\omega)$ for an intermediate wave number $q/q_\textnormal{F}\approx1.4$ for $N=34$ unpolarized electrons at a metallic density, $r_s=6$, for four different temperatures. Note that the line type (i.e., solid, dashed, etc.) distinguishes $\theta$, while the different colors correspond to different methods. More specifically, the green curves depict the random phase approximation (RPA), where exchange-correlation effects on the density-response are completely neglected, i.e., $G_\textnormal{RPA}(\mathbf{q},\omega) = 0$. Furthermore, the black curves have been obtained using the so-called static approximation, where the exchange-correlation effects are treated statically, $G_\textnormal{static}(\mathbf{q},\omega)=\textnormal{Re}G (\mathbf{q},0)$. 
For completeness, we note that here we use the exact static limit obtained from our PIMC simulations, Eq.~(\ref{eq:G_static}), which is in contrast to static dielectric theories like STLS or VS, cf.~Sec.~\ref{sec:static_response}.
Lastly, the red coloured curves correspond to the exact solutions that have been computed using our new stochastic sampling reconstruction method for the dynamic local field correction, which has been introduced in detail in the previous section. In addition, the shaded red area depicts the corresponding degree of uncertainty, which is estimated as the variance of all accepted trial solutions $S_\textnormal{trial}(\mathbf{q},\omega)$, Eq.~(\ref{eq:delta_S}).

At the largest considered temperature, $\theta=4$ (dotted curves), the influence of the LCF is relatively weak. Consequently, the uncertainty interval in the exact solution is relatively small. Furthermore, quantum effects do not dominate the dynamic density response, and the static approximation is in exact agreement with the DLFC curve over the entire $\omega$-range. In contrast, the RPA curve only qualitatively reproduces the other two data sets even at these relatively weakly coupled parameters.

Upon decreasing the temperature, the system simultaneously becomes more strongly coupled and more quantum mechanical. The first fact leads to severe systematic errors in the RPA, which significantly overestimates the peak position and underestimates the height. On the other hand, the quantum nature of the dynamic density response cannot be fully captured by a static local field correction, and the accuracy of the black curves declines with decreasing $\theta$. Still, we stress that the SLFC constitutes a substantial improvement over the RPA everywhere, and the observed disagreements are quantitative, but not qualitative.
Moreover, the impact of the frequency dependence of $G(\mathbf{q},\omega)$ is even less severe than in the previously investigated example of $r_s=10$, see Ref.~\cite{dornheim_dynamic}.

\begin{figure*}
\includegraphics[width=0.800647\textwidth]{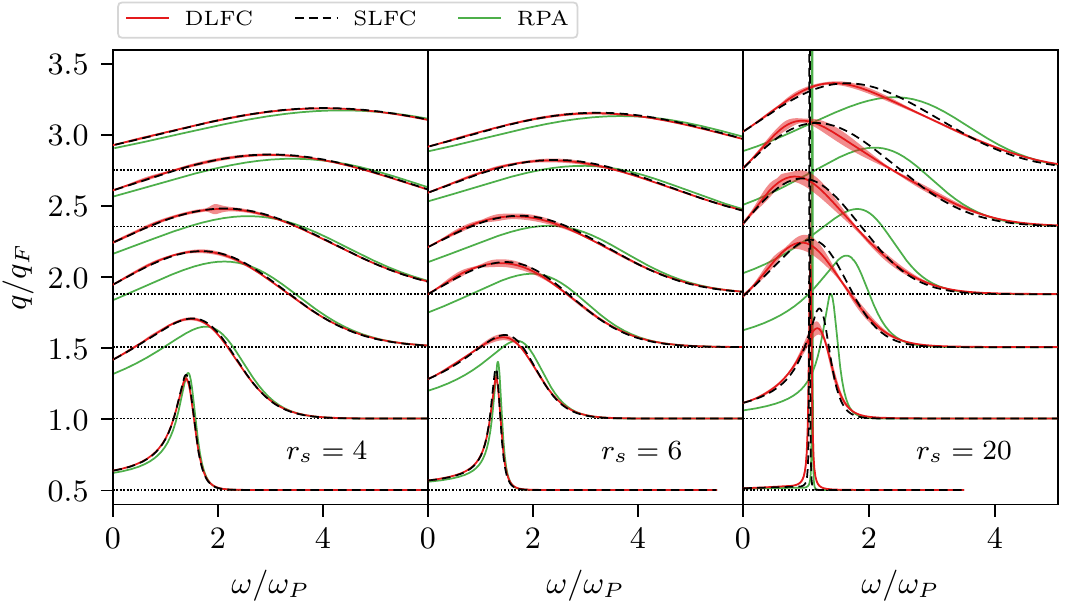}
\caption{\label{fig:breit}
Dispersion relation of the dynamic structure factor $S(\mathbf{q},\omega)$ of the warm dense electron gas with $N=66$ and $\theta=2$ for $r_s=4$ (left), $r_s=6$ (center), and $r_s=20$ (right). The solid red, dashed black, and solid green lines depict data from our stochastically sampled dynamic local field correction, the static approximation $G_\textnormal{static}(\mathbf{q},\omega)=\textnormal{Re}G(\mathbf{q},0)$, and the random phase approximation (RPA), respectively. The curves have been re-scaled for a better comparability.
}
\end{figure*}

Finally, we show entire dispersion relations of the dynamic structure factor for $N=66$ unpolarized electrons at $\theta=2$ for $r_s=4$, $r_s=6$, and $r_s=20$ in Fig.~\ref{fig:breit}.


First and foremost, we find that both at $r_s=4$
and $r_s=6$, which correspond to metallic densities falling well into the warm dense matter regime, the static approximation cannot be distinguished from the exact solution over the entire range of wave numbers $q$. In contrast, the RPA again only exhibits a qualitative agreement and is fairly inaccurate both regarding peak position and width for intermediate wave numbers, $q/q_\textnormal{F}\sim2$.
At $r_s=20$, which is a strongly coupled system more closely resembling an electron liquid, the total negligence of exchange-correlation effects renders the RPA completely inappropriate. The static approximation, on the other hand, constitutes a spectacular improvement and almost exactly reproduces the exact solutions found via our stochastic sampling scheme. In particular, the incipient excitonic feature, which manifests as a pronounced red-shift at intermediate wave numbers, is fully captured by the black curves~\cite{dornheim_dynamic}.


Therefore, we conclude that the static approximation [using our exact PIMC data for $G(q)$] does indeed provide a reliable and readily available tool to accurately compute the dynamic structure factor $S(\mathbf{q},\omega)$. This is particularly important at even lower temperatures and higher densities, where the present combination of PIMC and the stochastic sampling of the DLFC is prevented by the fermion sign problem, but $G_\textnormal{static}$ is accessible via the more advanced PB-PIMC and CPIMC methods~\cite{dornheim_pre,groth_jcp}.

\section{Summary and discussion\label{sec:summary}}

In this work, we have presented an \textit{ab initio} path integral Monte Carlo study of the density-response of the uniform electron gas at extreme temperature and density. 

First of all, we have introduced in detail the underlying theory, in particular the concept of the imaginary-time correlation-function $F(\mathbf{q},\tau)$, and its connection to the desired response functions. In particular, our new extensive PIMC data for $F$ have allowed us to present exact data for the static density-response function $\chi(\mathbf{q})$ and the corresponding static local field correction $G(\mathbf{q})$ from warm dense matter to the strongly correlated electron liquid regime. A comparison of these data to standard dielectric theories has revealed the following: (a) the random phase approximation reveals significant inaccuracies even at $r_s=4$ and $\theta=2$ and fully breaks down at $r_s=20$; (b) the approximate static local field correction by Vashista-Singwi leads to a significantly improved description of $\chi(\mathbf{q})$, and $G_\textnormal{VS}(\mathbf{q})$ is highly accurate at small $q$, which is due to the incorporation of the compressibility sum-rule. However, for larger wave numbers, the static LFC exhibits significant systematic deviations; (c) the static LFC due to Singwi-Tosi-Land-Sj\"olander constitutes the most accurate dielectric method both for $\chi$ and $G$, although at strong coupling ($r_s=20$) it, too, is only capable to provide a qualitative description of the static density-response. In this context, a future comparison of our exact PIMC data to the new hypernetted-chain (HNC) based static LFC by Tanaka~\cite{tanaka_hnc} and the quantum STLS (qSTLS) method~\cite{schweng,arora} might be valuable to evaluate the source of the errors in the dielectric methods: if the HNC approach would constitute a significant improvement at large $r_s$, the error in the STLS and VS schemes would be the approximate treatment of exchange-correlation effects in the static local field correction; a superior performance of qSTLS, on the other hand, would indicate the importance of the frequency dependence of $G(\mathbf{q},\omega)$, which is neglected in STLS, VS, and HNC, but consistently included in qSTLS.

Secondly, our study of the high-frequency limit of $G(\mathbf{q},\omega)$ at twice the Fermi temperature has revealed different behaviours at large wave numbers depending on the coupling parameter $r_s$: for small $r_s$, $G(\mathbf{q},\omega=\infty)$ increases with $q$, whereas the opposite holds at strong coupling (even at $r_s=6$) and $G(\mathbf{q},\omega=\infty)$ actually attains negative values. This is consistent with previous findings at zero temperature~\cite{iwamoto} and was explained by a nontrivial behaviour of the exchange--correlation part of the kinetic energy, $K_\textnormal{XC}$, shown in Fig.~\ref{fig:Kxc}.

Thirdly, we have given both an extensive introduction to the theoretical aspects and a practical demonstration of our new reconstruction method that allows us to obtain highly accurate results for the dynamic structure factor $S(\mathbf{q},\omega)$ purely on the basis of our PIMC data for the thermodynamic equilibrium, see Sec.~\ref{sec:stochastic_sampling_results}. While the basic idea and first results for $S(\mathbf{q},\omega)$ were already shown in Ref.~\cite{dornheim_dynamic}, here we provide an explicit demonstration of the stochastic sampling of the dynamic LFC, and how this leads to trial solutions $S_\textnormal{trial}(\mathbf{q},\omega)$, which are subsequently compared to our PIMC data for $F(\mathbf{q},\tau)$ and $\braket{\omega^k}$. We are confident that this comprehensive discussion will render our findings replicable and make the future adaption of our scheme to related problems in other fields possible.
In addition, the most significant new finding is the comparably high quality in the final results for the dynamic structure factor, which is in stark contrast to the real and imaginary parts of $G(\mathbf{q},\omega)$. More specifically, quite different trial solutions for $G_{\textnormal{trial},i}(\mathbf{q},\omega)$ lead to nearly identical $S_{\textnormal{trial},i}(\mathbf{q},\omega)$ and, therefore, cannot be selected or discarded on the basis of $F$ and $\braket{\omega^k}$, which explicitly depend on the dynamic structure factor, only.

In the fourth place, we have shown extensive new data for the dynamic structure factor of the UEG for hitherto unexplored parameters. Fully in accord with Ref.~\cite{dornheim_dynamic}, we find significant systematic errors in the random phase approximation even at relatively weak coupling, which are most pronounced around twice the Fermi wave number $q_\textnormal{F}$. In contrast, the static approximation, i.e., using the exact static limit of $G$ for all frequencies, leads to nearly exact results for even for the electron liquid at $r_s=20$. Furthermore, it retains an impressive accuracy with decreasing temperature and, in a nutshell, allows for a good description of $S(\mathbf{q},\omega)$ for all investigated parameters.

Let us conclude this paper by outlining a number of interesting topics for future research.

    Accurate data for $S(\mathbf{q},\omega)$ of the warm dense UEG are important for the interpretation of experiments within the widespread Chihara decomposition~\cite{chihara1,chihara2}. To this end, a readily available parametrization of the static local field correction $G(\mathbf{q};r_s,\theta)$ connecting the ground state results~\cite{cdop,farid} with PIMC results at finite temperature constitutes a highly desirable goal. Further, the PIMC data for the LFC could be complemented by configuration PIMC~\cite{groth_jcp} and permutation blocking PIMC data~\cite{dornheim_pre} at stronger quantum degeneracy (i.e., low temperature and high density), where our present strategy fails due to the fermion sign problem.

     Our new data for $S(\mathbf{q},\omega)$ can also be used to benchmark approximate methods for the description of quantum dynamics at finite temperature, such as nonequilibrium Green functions~\cite{kas1,kas2,kas3,kwong} and the method of moments by Tkachenko and co-workers~\cite{igor1,igor2,igor_quantum}.

    Moreover, the observed negative dispersion relation in $S(\mathbf{q},\omega)$ (see also Ref.~\cite{dornheim_dynamic}) constitutes an interesting feature in its own right. While the interpretation as an incipient excitonic feature from Refs.~\cite{takada1,takada2,higuchi} is plausible, its manifestation in the warm dense UEG deserves a more detailed investigation. For completeness, we mention that a similar behaviour has been observed in the classical one-component plasma~\cite{ott}, and in experiments with alkali metals~\cite{vomFelde}.

    Finally, the stochastic sampling method for the reconstruction of a dynamic quantity is, in principle, not limited to the present application to PIMC data for the UEG. For example, while the computation of imaginary-time correlation-functions in ground state QMC simulations~\cite{wmc_review} of fermions is nontrivial~\cite{motta1}, Motta \textit{et al.}~\cite{motta2} recently obtained accurate results for $F(\mathbf{q},\tau)$ for a $2D$ UEG at zero temperature using a phaseless auxiliary field QMC approach.

 \section*{Acknowledgments}
 S.G.~and T.D.~contributed equally to this work.
 We acknowledge T.~Sjostrom for providing the VS data shown in Figs.~\ref{fig:CHI_theta2} and \ref{fig:LFC_theta2}, and fruitful discussions with H.~K\"ahlert, Zh.A.~Moldabekov, and M.~Bonitz.
This work has been supported by the Deutsche Forschungsgemeinschaft via project BO1366 and by the Norddeutscher Verbund f\"ur Hoch- und H\"ochleistungsrechnen (HLRN) via grant shp00015 for CPU time.

\end{document}